\documentclass[acmsmall,table,xcdraw]{acmart}

\usepackage{graphicx}
\usepackage{soul} 
\usepackage{soul} 
\usepackage{paralist} 
\usepackage{listings}
\usepackage{tikz}
\usepackage{tikz}
\usepackage{pgfplots}
\usepackage{pgfplotstable}
\pgfplotsset{compat=1.17}

\usetikzlibrary{positioning, arrows.meta, shapes, calc}

\usepackage{listings}
\usepackage{xcolor}

\lstdefinestyle{mystyle}{
  backgroundcolor=\color{gray!5},
  basicstyle=\ttfamily\small,
  breaklines=true,
  frame=single,
  framerule=0.3pt,
  columns=fullflexible,
  keepspaces=true,
  tabsize=2,
  showstringspaces=false,
  captionpos=b
}
\usepackage{xspace} 
\usepackage{subcaption} 
\usepackage{multirow} 
\usepackage{siunitx} 
\usepackage[many]{tcolorbox} 
\usepackage{adjustbox}
\usepackage{tikz}
\usetikzlibrary{arrows.meta,positioning,calc,shapes}

\usepackage{booktabs}
\usepackage{lscape} 
\usepackage{graphicx}
\usepackage{adjustbox}
\usepackage{booktabs}
\usepackage{tikz}
\usetikzlibrary{arrows.meta, positioning, shapes.geometric, fit, backgrounds}
\usepackage{xurl} 
\usepackage{ifthen} 

\usepackage[utf8]{inputenc}
\usepackage[T1]{fontenc}
\usepackage{microtype}
\usepackage{tcolorbox}
\usepackage{xcolor}
\usepackage{listings}
\usepackage{caption}
\usepackage{enumitem}

\acmJournal{TOSEM}
\acmYear{2025}
\acmVolume{0}
\acmNumber{0}
\acmArticle{0}
\acmMonth{10}

\definecolor{codebg}{RGB}{250,250,250}
\definecolor{boxblue}{RGB}{240,248,255}
\definecolor{boxborder}{RGB}{0,102,204}
\definecolor{keyword}{RGB}{127,0,85}
\definecolor{commentcolor}{RGB}{63,127,95}
\definecolor{stringcolor}{RGB}{42,0,255}

\tcbset{
  myboxstyle/.style={
    enhanced,
    breakable,
    colback=boxblue,
    colframe=boxborder,
    arc=2mm,
    boxrule=0.8pt,
    left=4pt,
    right=4pt,
    top=4pt,
    bottom=4pt,
    width=\textwidth,     
    enlarge left by=0mm,
    overlay={
      \begin{tcbclipinterior}
        \fill[white,opacity=0] (interior.south west) rectangle (interior.north east);
      \end{tcbclipinterior}
    }
  }
}

\lstdefinestyle{customjava}{
  language=Java,
  basicstyle=\ttfamily\small,
  backgroundcolor=\color{codebg},
  keywordstyle=\color{keyword}\bfseries,
  commentstyle=\color{commentcolor}\itshape,
  stringstyle=\color{stringcolor},
  showstringspaces=false,
  breaklines=true,
  frame=single,
  framerule=0.4pt,
  rulecolor=\color{gray!40},
  xleftmargin=3pt,
  xrightmargin=3pt
}
\lstdefinestyle{custompython}{
  language=Python,
  basicstyle=\ttfamily\small,
  backgroundcolor=\color{codebg},
  keywordstyle=\color{keyword}\bfseries,
  commentstyle=\color{commentcolor}\itshape,
  stringstyle=\color{stringcolor},
  showstringspaces=false,
  breaklines=true,
  frame=single,
  framerule=0.4pt,
  rulecolor=\color{gray!40},
  xleftmargin=3pt,
  xrightmargin=3pt
}

\setlist[itemize]{leftmargin=*,topsep=2pt,partopsep=0pt,parsep=0pt,itemsep=2pt}

\usepackage{tabularx}
\usepackage{multirow}
\usepackage{array}
\usepackage{caption}
\usepackage{graphicx}
\usepackage{siunitx}   
\usepackage{pgf}       

\usepackage{tikz}
\usetikzlibrary{arrows.meta, positioning, shapes.geometric, backgrounds, fit}

\newcolumntype{Y}{>{\raggedright\arraybackslash}X}

\AtBeginDocument{%
  \providecommand\BibTeX{{%
    \normalfont B\kern-0.5em{\scshape i\kern-0.25em b}\kern-0.8em\TeX}}}

\acmJournal{TOSEM}

\newtcolorbox{mybox}[2][]{
top=0.15in,left=4pt,right=4pt,bottom=4pt,
fonttitle=\bfseries,
colbacktitle=gray,
colback=gray!5,
colframe=gray!40!black,
enhanced,
attach boxed title to top left={xshift=1.5em,yshift=-\tcboxedtitleheight/2},
boxed title style={size=small},
drop shadow={black!50!white},
title=#2,#1}

\newcommand{\countobservations}{
    \def \countobservations{1}
}
\newcounter{observation}
\countobservations

\newcommand{\countimplications}{
    \def \countimplications{1}
}
\newcounter{implication}
\countimplications

\newboolean{showcomments}
\setboolean{showcomments}{true}

\ifthenelse{\boolean{showcomments}}
{\newcommand{\nbc}[3]{
 {\colorbox{#3}{\bfseries\sffamily\scriptsize\textcolor{white}{#1}}}
 {\textcolor{#3}{\sf\small$\blacktriangleright$\textit{#2}$\blacktriangleleft$}}
 }
}
{\newcommand{\nbc}[3]{}
 }

\graphicspath{{figures/}}

\begin{document}

\title{RAG-Reflect: Agentic Retrieval-Augmented Generation with Reflections for Comment-Driven Code Maintenance on Stack Overflow}
.
\author{Mehedi Hasan Shanto}
\email{shanto1@uwindsor.ca}
\affiliation{%
  \institution{School of Computer Science at University of Windsor}
  \city{Ontario}
  \country{Canada}
}
\author{Muhammad Asaduzzaman}
\email{md.asaduzzaman@queensu.ca}
\affiliation{%
  \institution{School of Computer Science at University of Windsor}
  \city{Ontario}
  \country{Canada}
}
\author{Alioune Ngom}
\email{angom@uwindsor.ca}
\affiliation{%
  \institution{School of Computer Science at University of Windsor}
  \city{Ontario}
  \country{Canada}
}

\renewcommand{\shortauthors}{Shanto, et al.}

\begin{abstract}
  
User comments on online programming platforms such as Stack Overflow play a vital role in maintaining the correctness and relevance of shared code examples. However, the majority of comments express gratitude or clarification, while only a small fraction highlight actionable issues that drive meaningful edits. Automatically identifying which comments genuinely motivate code changes is critical for sustaining knowledge quality, automating post updates, and mitigating information decay in developer forums. This paper demonstrates how agentic AI principles can revolutionize software maintenance tasks by presenting \textbf{RAG-Reflect}, a modular framework that achieves fine-tuned-level performance for valid comment-edit prediction without task-specific training. \textbf{Valid Comment--Edit Prediction (VCP)} task---determining whether a user comment directly triggered a subsequent code edit. The framework integrates large language models (LLMs) with retrieval-augmented reasoning and self-reflection mechanisms. \textbf{RAG-Reflect} operates through a three-stage runtime workflow built on a one-time pattern analysis phase. During initialization, an Interpretation module analyzes the knowledge base to generate validation rules. At inference time, the system (1) retrieves contextual examples, (2) reasons about comment-edit causality, and (3) reflects on decisions using the pre-established rules. We evaluate RAG-Reflect on the publicly available \textbf{SOUP} benchmark, achieving \textbf{Precision = 0.81}, \textbf{Recall = 0.74}, and \textbf{F1 = 0.78}---outperforming traditional baselines (e.g., Logistic Regression, XGBoost, different prompting techniques) and closely approaching the performance of fine-tuned models (\textit{F1 = 0.773}) without retraining. Our ablation and stage-level analyses show that both retrieval and reflection modules substantially enhance performance. By decomposing perception, memory, reasoning, and reflection into modular components, RAG-Reflect demonstrates that agentic reasoning can achieve fine-tuned-level accuracy while remaining lightweight, extensible, and adaptable. These findings offer new insights into designing agentic LLM workflows for software engineering and open a path toward self-improving, data-efficient reasoning systems.

\end{abstract}



\maketitle
\section{Introduction}

Stack Overflow (SO) is one of the most widely used platforms for software developers to seek and share programming knowledge. \textbf{Jeff Atwood}, the co-founder of SO, emphasized that the goal of SO is not merely to answer individual questions, but to \textbf{“collaboratively build an artifact that will benefit future coders”} – a vision that depends on continually maintaining and improving posts over time \cite{SHEIKHAEI2023111590}. Over the years, developers have increasingly relied on SO not only to resolve specific bugs, but also to copy, adapt, and reuse code snippets directly in real-world projects~\cite{Wu2019UtilizeSOCode,Zhang2021ReadingAnswersNotEnough}. Prior work shows that SO content strongly influences how developers design and implement their systems, and that many questions and answers are revisited long after their initial posting~\cite{Rahman2015Unresolved,Wang2013SACInteractions,Baltes2018}. Consequently, the long-term quality, correctness, and freshness of SO answers directly affect the reliability and maintainability of software that depends on them~\cite{Zhang2021ObsoleteAnswers}.

\begin{figure}[ht]
\centering
\includegraphics[width=0.95\linewidth]{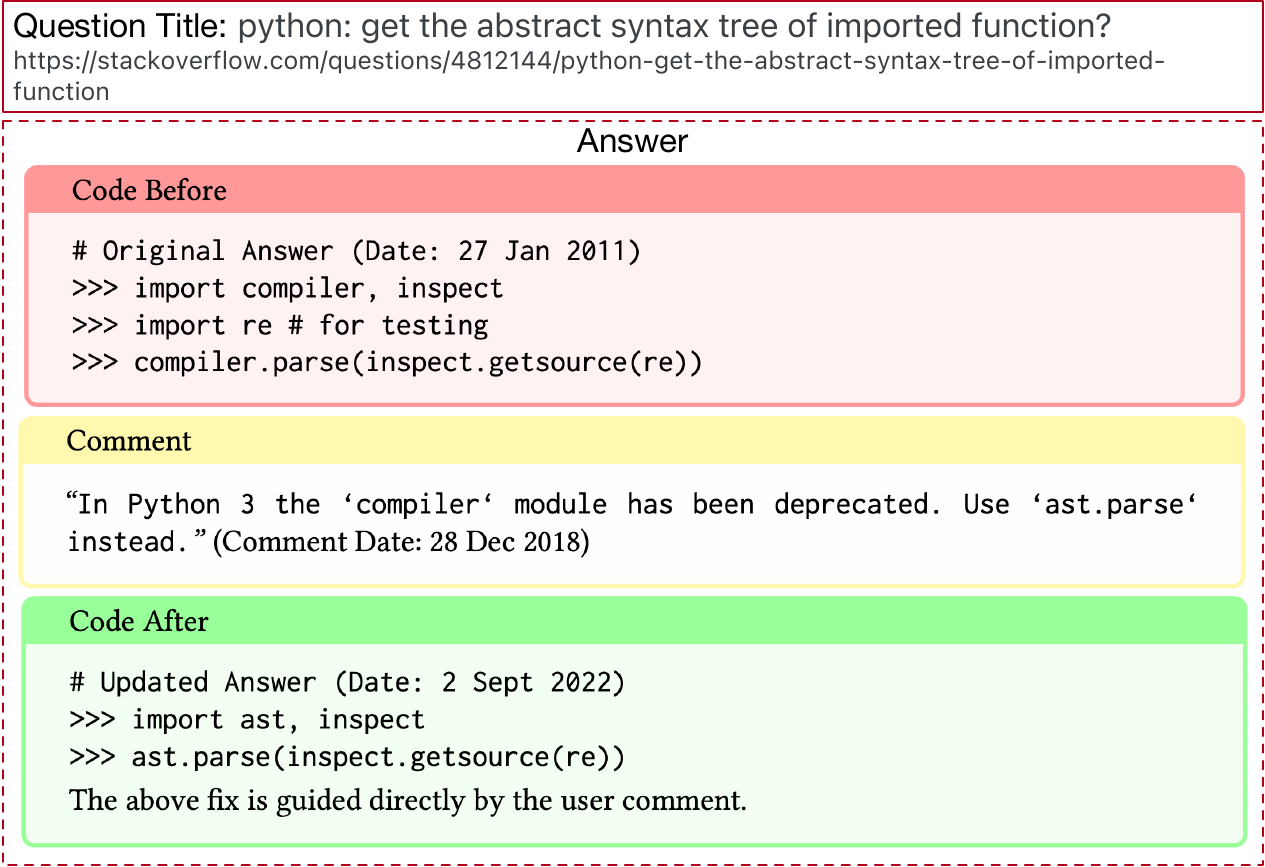}
\caption{An example of a comment that triggered the edit to a Stack Overflow answer.}
\Description{}
\label{fig:outdated-api-example}
\end{figure}

Stack Overflow allows its users to comment on a post (i.e., either a question or an answer). Comments often capture exactly the kind of time-sensitive knowledge that must be incorporated into answers. Consider the example in Fig.~\ref{fig:outdated-api-example}. An older Python answer uses the \texttt{compiler.parse()} to get the abstract syntax tree of imported function. However, the compiler module is deprecated in the newer version of Python (i.e., Python 3). A user suggested to use \texttt{ast.parse} instead of \texttt{compiler.parse} by adding a comment to the answer. This concise remark provides three essential signals: (i) \emph{why} the existing answer is now incorrect, (ii) \emph{what} the correct fix is, and (iii) \emph{how} to update the code. After reading this comment, an editor (or an automated system) can revise the snippet by replacing \texttt{compiler.parse} with \texttt{ast.parse}, restoring correctness for future readers. 

Although comments frequently identify real issues, most comments do \emph{not} warrant edits. Many are conversational, vague, opinion-based, or unrelated to the code. Therefore, an essential prerequisite to any automated post-maintenance approach is the ability to determine whether a given comment should actually lead to an edit. We refer to this as the \textbf{Valid Comment--Edit Pair (VCP)} task. A comment--edit pair is valid if (i) the comment is relevant to the answer, (ii) it identifies a real issue or improvement, and (iii) the corresponding edit directly addresses that comment without introducing unrelated changes. Identifying valid comment-edit pairs is crucial because such pairs help users to understand the rationale for code changes. Comments therefore act as lightweight, conversational “micro-reviews” that can flag obsolete APIs, security vulnerabilities, incorrect assumptions, or missing edge cases in answers~\cite{Zhang2021CommentsOrganized,Wang2013SACInteractions}. In practice, however, this loop is often broken. Empirical studies show that many answers become obsolete without being updated~\cite{Zhang2021ObsoleteAnswers}, and that comments frequently request improvements which never materialize in the corresponding posts~\cite{SHEIKHAEI2023111590,Soni2019}. Sheikhaei et al.\ report that a substantial proportion of \emph{update request comments} in SO answers remain unaddressed even after long periods of time~\cite{SHEIKHAEI2023111590}, while Soni and Nadi find that comment-induced updates are relatively rare and often delayed~\cite{Soni2019}. As a result, important technical insights remain buried in the comment section, which is partially hidden by default and harder to scan than the top-voted answer~\cite{Zhang2021CommentsOrganized,Zhang2021ReadingAnswersNotEnough}. This misalignment between where corrections appear (comments) and where most users look (answers) leads to a knowledge maintenance problem: developers can be misled by outdated or incomplete answers even when the correct information has already been discussed in comments~\cite{Zhang2021ObsoleteAnswers}.

This gap has motivated a growing body of work on \emph{automatic post updating} and collaborative editing assistance in Q\&A communities. Chen et al.\ propose deep learning techniques to support collaborative editing on Q\&A sites, helping the community refine and improve posts~\cite{Chen2017CollaborativeEditing}. Jin and Servant study the types of edits performed on highly answered SO questions~\cite{Jin2019HighlyAnsweredEdits}, while Li et al.\ and others analyze the trade-offs of importing Wikipedia-style collaborative editing mechanisms into Q\&A platforms~\cite{Li2015Wikipedia,Kittur2008Crowds,Hu2007WikiQuality}.  Focusing specifically on the link between comments and edits, Tang and Nadi mine SO’s revision history to build a dataset of answer \emph{comment--edit pairs} and use it to recommend code maintenance changes~\cite{Tang2021}. Their approach relies on identifying cases where a comment is followed by an edit to the same answer and treating such pairs as supervision for learning to recommend similar edits. Soni and Nadi analyze comment-induced updates and characterize when and how comments trigger edits~\cite{Soni2019}. More recently, Mai et al.\ introduce the SOUP (Stack Overflow Post Updating) framework, which formulates the problem as two tasks: (1) Valid Comment--Edit Prediction (VCP), and (2) Automatic Post Updating (APU)~\cite{Mai2025SOUP}. SOUP fine-tunes a large language model (LLM), Code Llama~\cite{Roziere2023}, on a curated dataset of 78k Java comment--edit pairs, achieving notable performance both on automatic metrics and in-the-wild submissions of generated edits to SO~\cite{Mai2025SOUP}. These works demonstrate that learning from historical comment--edit pairs can help automate post maintenance, and that LLMs are powerful engines for code-aware editing~\cite{Brown2020GPT3,Chen2021Codex,touvron2023llama,zhao2023survey}.

Despite these advances, existing approaches suffer from several limitations. First, they often depend on supervised fine-tuning over historically edited examples, which are inherently biased: the vast majority of issues raised in comments were never fixed and thus never appear in the training data~\cite{SHEIKHAEI2023111590,Zhang2021ObsoleteAnswers}. This leads to low recall of valid comment--edit relationships and constrains the learned models to imitate a narrow set of edits~\cite{Tang2021,Mai2025SOUP}. Second, models trained on a particular language (e.g., Java) or on specific update types may generalize poorly to other languages, domains, or more complex reasoning-intensive changes without additional expensive fine-tuning~\cite{Ren2020CodeBLEU,Chen2021Codex}. Third, most existing systems treat comment-driven updating as a single-step transformation: given the pre-edit answer and the comment, the model directly predicts the post-edit answer~\cite{Tang2021,Mai2025SOUP}. Such one-shot generation is brittle in knowledge-intensive, multi-hop scenarios where a good update requires fetching external documentation, reconciling multiple comments, or verifying that the revised code still compiles and behaves correctly~\cite{Lewis2020,Parvez2021}. Finally, there is a cost and agility concern: large-scale fine-tuning for every new domain or task variant is expensive and does not adapt quickly to emerging APIs or evolving community norms~\cite{zhao2023survey}. 

Recent work on \emph{agentic AI} and \emph{LLM-based agents} suggests an alternative path. Instead of relying solely on monolithic fine-tuned models, agentic systems orchestrate LLMs as decision-making agents that can plan, use tools (e.g., search, code execution), and iteratively reflect on their outputs~\cite{Acharya2025Survey}. Techniques such as chain-of-thought reasoning~\cite{Wei2022CoT}, Tree-of-Thoughts~\cite{Yao2023TreeOfThought}, and reflective self-improvement~\cite{Shinn2023Reflexion,asai2023selfRAG,Jiang2023} have shown that decomposing tasks into multi-step workflows often yields more reliable and interpretable behavior than single-shot prompting. In the context of code and software engineering, autonomous and semi-autonomous agents have been successfully applied to program repair, debugging, and multi-step coding tasks~\cite{Zhang2024AutoCodeRover,Yang2024SWEAgent,Roychoudhury2025,Li2015Wikipedia}. Ng’s recent guidelines for building agentic AI workflows emphasize that, for many practical applications, carefully designed multi-step workflows with retrieval and reflection provide better returns than aggressively fine-tuning larger models for one-step use. As he has stated, \textbf{“\textit{So I tend not to fine-tune a model until I've really exhausted the other options, because fine-tuning tends to be quite complex.}”}~\cite{Ng2024AgenticWorkflow}. This perspective aligns with emerging visions in software engineering that propose agentic AI as a foundation for future development workflows~\cite{Roychoudhury2025,Hassan2025Roadmap,Li2015AnswerQuality,Akbar2025Practitioner}.

Motivated by these insights, we propose \textbf{RAG-Reflect}, a novel framework that combines \emph{Retrieval-Augmented Generation} (RAG)~\cite{Lewis2020,Parvez2021} with an \emph{agentic, reflective workflow} to automate the comment-to-edit loop for Stack Overflow answers. Rather than modeling post updating as a single sequence-to-sequence mapping, RAG-Reflect reframes comment-driven maintenance as a multi-step, agentic process. For a given post and its comments, the system first retrieves semantically similar historical comment-edit pairs to provide contextual grounding. It then reasons over this context to predict whether a comment validly motivated an edit (VCP). Concurrently, a one-time interpretation of the knowledge base generates a rule-based prompt that captures distinguishing patterns between valid and invalid pairs. Finally, the reflection module applies these rules to critique and refine the initial prediction, ensuring the final decision is precise, consistent, and contextually justified. Our workflow design is inspired by recent agentic RAG patterns, where retrieval, generation, and critique are tightly integrated in a loop~\cite{asai2023selfRAG,Singh2025x,Panda2025AgenticRAG,Xu2025HumanRetrieval}. By grounding edits in retrieved context and explicitly checking alignment between the comment, the original code, and the proposed changes, RAG-Reflect aims to produce updates that (a) faithfully address the issues raised in comments, (b) avoid unrelated or overly invasive modifications, and (c) generalize across languages and update types without task-specific fine-tuning.

We evaluate RAG-Reflect on a curated dataset of Stack Overflow answers, comments, and edits, and compare it against prior approaches including heuristic comment--edit matching~\cite{Tang2021}, fine-tuned LLM baselines~\cite{Mai2025SOUP}, and non-agentic RAG pipelines~\cite{Lewis2020,Parvez2021}. Our results show that RAG-Reflect improves the performance of predicting valid comment-edit pairs, and yields higher-quality post updates. Furthermore, ablation studies demonstrate that both retrieval and reflection components contribute significantly to the performance, underscoring the value of an agentic design for this task.

In summary, this paper makes the following contributions:
\begin{itemize}
  \item We introduce \textbf{RAG-Reflect}, an agentic RAG-based framework that automates the comment-to-edit workflow for Stack Overflow answers, integrating retrieval, planning, and reflection to produce high-quality updates.
  \item We propose an autonomous code-update pipeline that generates comment-aligned edits while constraining unnecessary changes, drawing on recent advances in LLM-based agents and reflective improvement~\cite{Shinn2023Reflexion,asai2023selfRAG,Acharya2025Survey,Ng2024AgenticWorkflow}.
  \item We conduct a comprehensive empirical evaluation of \textbf{RAG-Reflect} against state-of-the-art techniques for both predicting valid-comment pairs (VCP) and automatic post update (APU) tasks.

  \end{itemize}

The remainder of this paper is organized as follows. 
Section~\ref{sec:background} describes the background of stack overflow.
Section~\ref{motivation} motivates for the study.
Section~\ref{sec:relatedwork} reviews the existing literature on stack overflow comment–code consistency, and agentic LLM workflows.
Section~\ref{sec:agentic-workflow} introduces our proposed RAG-Reflect framework. 
Section~\ref{sec:evaluation} presents the benchmark datasets, evaluation metrics used in our study. 
Section~\ref{sec:results} reports the empirical findings and addresses each of the research questions in turn. 
Section~\ref{sec:discussion} discusses the broader implications of our results for researchers, practitioners, and tool designers. 
Section~\ref{sec:threats_to_validity} outlines potential threats to validity. 
Finally, Section~\ref{conclusion} concludes the paper and suggests directions for future work.

\section{Background}
\label{sec:background}
In this section, we briefly overview the question answering process in SO, and discuss how developers edit a question or an answer.

\begin{figure}[ht]
\centering
\includegraphics[width=0.9\linewidth]{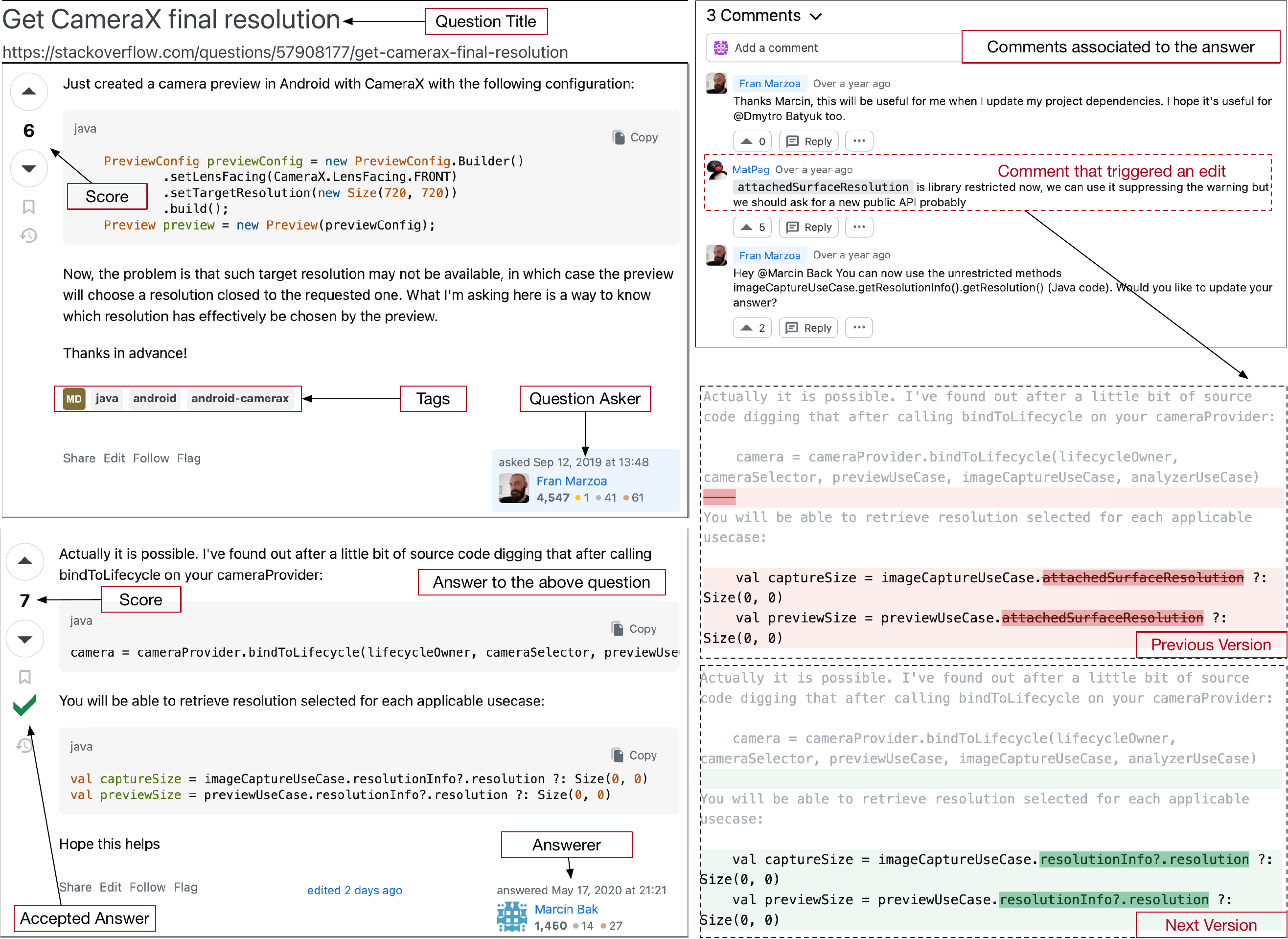}
\caption{An example of a question and the accepted answer in Stack Overflow. A comment associated to the answer triggered an edit performed by the question asker to maintain the freshness of the answer.}
\Description{A screenshot showing a Stack Overflow question, the accepted answer, and a comment that prompted an edit to the answer.}
\label{fig:question_answering_stackoverflow}
\end{figure}

Stack Overflow encourages its registered users to ask any programming-related questions. Each question consists of a title, a body, and a set of tags to identify the topics of that question. Each question can receive one or more answers. The author of the question can select one of the answer as the accepted answer that completely solve the problem in question. Figure~\ref{fig:question_answering_stackoverflow} shows an example of a question and the associated answer. Each question and answer can receive upvotes or downvotes. Voting provides a mechanism to separate poor-quality questions or answers from the good ones. The difference between the number of upvotes and the number of downvotes received by a post is referred to as the post's score. Stack Overflow users obtain points by participating in community activities (i.e., asking good questions, providing useful answers), known as reputation points. The reputation indicates the degree of a user's contribution to SO and is a rough measurement of how much the community trusts the user\footnote{\url{https://stackoverflow.com/help/whats-reputation}}.

Both questions and answers can receive one or more comments. The commenting mechanism plays a vital role in this ecosystem by enabling lightweight, reputation-neutral feedback. Comments allow users to request clarification, point out inaccuracies, flag outdated or deprecated APIs, or suggest improvements without modifying the original post~\cite{Anderson2013Badges, Faruqui2018WellFormed}. Unlike posts, comments cannot receive downvotes and do not affect a user’s reputation; they can only be upvoted to increase visibility. This reinforces their purpose as transient, discussion-oriented contributions intended to support and refine the main content rather than serve as permanent components of the knowledge base.

Stack Overflow allows its users to edit a question or an answer to improve its quality. Authors can edit their own posts, and any other users who have more than 2000 reputation points can edit posts of other users. Registered users who do not have the required 2000 reputation points can suggest an edit, but those edits need to be reviewed by SO moderators before being accepted by the moderators ~\cite{Chen2017CollaborativeEditing, Kalliamvakou2015InnerSource}. Each post is version-controlled, meaning Stack Overflow not only preserves the edit or changes that happened to a post but also preserves the time of the edit, who performed the edit, and where the edit is performed (i.e., title, body, or tags)~\cite{Sabel2007RevisionHistory}.

\section{Motivating Examples: Implicit Comment Cues and Small Semantic Edits}
\label{motivation}

Prior work shows that, although many comments request or imply improvements, only a small fraction of them ever lead to edits on the corresponding answers~\cite{SHEIKHAEI2023111590,Soni2019,Mondal2023RejectedEdits}. A key reason is that these comments are often short, context-dependent, and only implicitly specify the desired change. Accurately linking such comments to edits requires understanding subtle semantic relationships between the comment text and the surrounding code, rather than relying on surface-level lexical overlap.

\begin{figure}[t]
\centering
\includegraphics[width=0.98\linewidth]{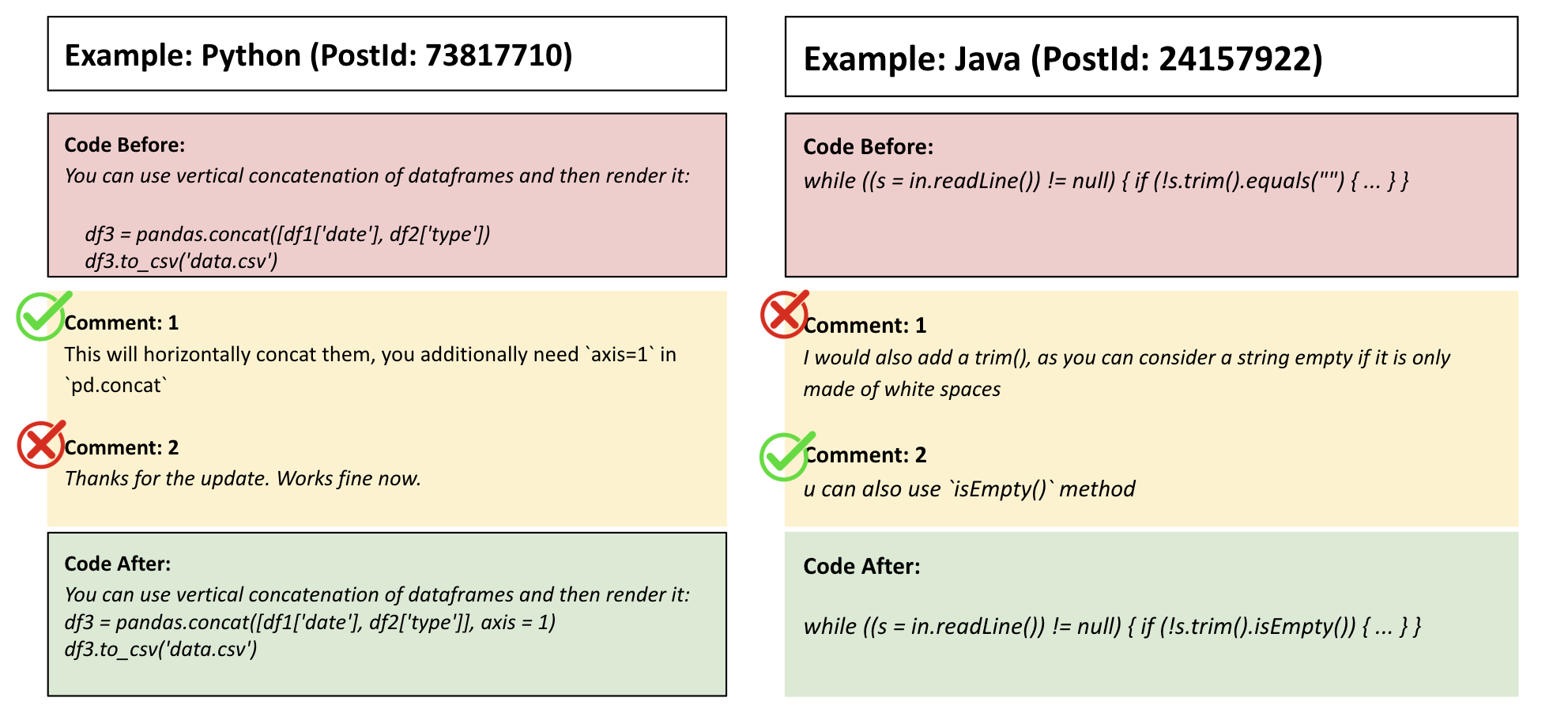}
\caption{Motivating examples of comment-driven edits on Stack Overflow. \textbf{Left}: a Python post where a comment implicitly indicates the need for \texttt{axis=1} in \texttt{pandas.concat}. \textbf{Right}: a Java post where a comment suggests the more idiomatic \texttt{isEmpty()} check. In both cases, small edits driven by concise comments are essential for correctness and code quality.}
\Description{}
\label{fig:motivation-examples}
\end{figure}

Figure~\ref{fig:motivation-examples} illustrates two typical examples drawn from SO. On the left, a Python answer suggests using
\texttt{pandas.concat} and provides a code snippet that vertically concatenates two columns, but omits the \texttt{axis=1} argument. A subsequent comment notes that ``this will horizontally concat them, you additionally need \texttt{axis=1} in \texttt{pd.concat}``. The revised code only differs by a few characters, yet that tiny edit is crucial for correctness of the code: without it, the snippet does not perform the operation described in the prose. On the right, a Java answer filters out empty lines using a compound condition \texttt{!s.trim().equals("")}. A commenter suggests using the more idiomatic \texttt{isEmpty()} method; the updated code replaces the equality check with \texttt{!s.trim().isEmpty()}, preserving behavior while improving readability and robustness.

These cases highlight three core challenges. First, the comments do not explicitly say ``this edit fixes bug X'' or ``replace line Y with Z''; instead, they provide hints that must be interpreted in the context of the existing code. Second, the resulting edits are small and localized, yet highly semantic: a single missing argument or method call can change correctness or clarity. Third, changes suggested in comments are language and API-specific (e.g., \texttt{axis} in Pandas, \texttt{isEmpty()} in Java), making it difficult for techniques that depend on lexical heuristics or shallow classifiers to generalize across posts and languages. 

Together, these examples motivate our design of \textbf{RAG-Reflect}. To determine whether a comment truly justifies an edit and to synthesize the corresponding change, an automated system must (i) reason about the intent of the comment, (ii) understand the original code’s behavior, and (iii) propose a minimal, targeted modification that aligns both with the comment and with language/library idioms. This need for contextual, multi-step reasoning is precisely where an agentic, retrieval-augmented, and reflective workflow can perform the best.

\section{Related Work}
\label{sec:relatedwork}

\subsection{Stack Overflow Comments and Post Edits}

The interplay between user comments and post updates on Stack Overflow (SO) has attracted significant attention in empirical software engineering research. Early studies identified that comments are not merely conversational artifacts but serve as *feedback mechanisms* that can influence code and textual revisions. 
Soni and Nadi~\cite{Soni2019} conducted one of the first systematic investigations into comment–edit relationships, introducing the concept of *comment-induced updates*. 
Their analysis revealed that while many comments on answers request improvements (e.g., pointing out errors, suggesting clarifications), only a small fraction (\textasciitilde5\%) actually lead to corresponding edits in the post. 
This observation highlighted a major inefficiency in community-driven knowledge maintenance: valuable suggestions often remain unincorporated.

Building on this foundation, Tang and Nadi~\cite{Tang2021} explored how comment–edit pairs can be leveraged to recommend code maintenance changes. 
By mining large-scale historical SO data, they extracted comment-edit relationships that captured real developer feedback, demonstrating their utility for automated recommendation models. 
Their study showed that most actionable comments address concrete correctness issues or deprecations and that a small subset of these pairs can be applied as real code improvements across projects.

Sheikhaei \textit{et al.}~\cite{SHEIKHAEI2023111590} expanded this analysis through a deep empirical study of *update request comments* (URCs)---comments that explicitly or implicitly request content updates.
From 1,221 manually annotated comments, they found that about half were URCs, and roughly two-thirds of them were addressed, typically through either an edit or a clarifying comment.
However, over 30\% of URCs remained unresolved even after a year, revealing a persistent gap between feedback recognition and code correction. 
Their classifier for URC detection achieved over 90\% accuracy, suggesting that linguistic cues (imperatives, negations, or references to outdated entities) are strong indicators of actionable feedback.

Complementary to this line of work, Mondal \textit{et al.}~\cite{Mondal2023RejectedEdits} analyzed the quality of edit proposals and introduced a model for predicting whether a suggested edit would be rejected by moderators. 
Wang et al. \cite{Wang2018} identified common rejection reasons, such as deviations from author intent or minor cosmetic changes in edit rejection prediction. 
Their subsequent tool, \textit{EditEx}, integrated real-time quality feedback into the SO editor, reducing edit rejections significantly in a user study.

Complementing these empirical studies, Mai \textit{et al.} introduced the SOUP task and dataset for \emph{automated Stack Overflow post updating}~\cite{Mai2025SOUP}. 
SOUP frames post maintenance as learning from historical comment–edit pairs to recommend concrete updates that improve answer quality. 
Their pipeline mines aligned \texttt{(comment, code\_before, code\_after)} triples and evaluates models that generate or select edits consistent with the comment’s intent. 
Results show that leveraging comment cues and historical edit analogies materially improves update quality—reinforcing the premise that actionable comments can be identified and operationalized for maintenance. 
In our work, we use SOUP strictly as an \emph{evaluation benchmark} for VCP, while our method remains dataset-agnostic and centers on agentic retrieval and reflection rather than fine-tuning.

Together, these studies demonstrate that comments are powerful but underutilized signals for maintaining SO’s knowledge base. 
Detecting which comments truly motivate edits (the Valid Comment–Edit Prediction problem) remains an open challenge, directly motivating our work.

\subsection{LLMs in Software Engineering Tasks}

Recent advances in large language models (LLMs) have led to frameworks that integrate retrieval-based augmentation and reasoning rather than relying purely on fine-tuning. 
Lewis \textit{et al.}~\cite{Lewis2020} introduced \textbf{Retrieval-Augmented Generation (RAG)}, where an encoder retrieves relevant documents from a large corpus before text generation, improving factual accuracy and grounding. 
This idea has been applied to software engineering tasks as well: Parvez \textit{et al.}~\cite{Parvez2021} proposed \textbf{RedCoder}, a retrieval-augmented model for code generation and summarization, which uses semantically similar examples as contextual input to improve generated code accuracy.

Beyond retrieval, \textbf{zero-shot} and \textbf{few-shot} reasoning approaches exploit in-context examples to adapt to new tasks without retraining. 
For example, Jiang \textit{et al.}~\cite{Jiang2023} demonstrated that large models can iteratively evolve code by self-prompting in multi-turn reasoning cycles. 
These works collectively support the premise that retrieval-based grounding combined with in-context reasoning can match or exceed task-specific fine-tuning performance while remaining data-agnostic.

Large language models trained on code---such as \textbf{Codex}~\cite{Chen2021Codex}, \textbf{CodeT5}~\cite{Wang2021CodeT5}, and \textbf{Code Llama}~\cite{Roziere2023}---have revolutionized code synthesis and summarization. 
However, their capability to reason about existing code and perform meaningful updates remains limited. 
Tufano \textit{et al.}~\cite{Tufano2019} pioneered learning meaningful code changes via neural machine translation, demonstrating that while models can learn common repair patterns, they struggle with complex or context-dependent edits.
Chen \textit{et al.}~\cite{Chen2021Codex} similarly reported that Codex, though powerful for code generation, fails on multi-step reasoning tasks such as error correction or dependency updates.

Recent developments have focused on \textbf{autonomous code agents} that iteratively refine their outputs through tool use and feedback. 
Yang \textit{et al.}~\cite{Yang2024SWEAgent} introduced \textbf{SWE-Agent}, an agent–computer interface that allows GPT-4 to edit code repositories using a sequence of structured commands (e.g., open file, modify line, run test). 
By integrating external tools (compiler, test suite) and iterative verification, SWE-Agent achieved over 87\% success in HumanEvalFix tasks—far exceeding non-agentic baselines. 
Similar frameworks that couple reflection with execution feedback (e.g., self-debugging and auto-repair agents) demonstrate that reflective reasoning is crucial for consistent improvement in code maintenance accuracy.

Andrew Ng~\cite{Ng2024AgenticWorkflow} argues that such modular strategies---combining retrieval, reasoning, and reflection---offer practical advantages over costly fine-tuning. 
As he states, This philosophy underpins our approach: rather than fine-tuning, we structure the workflow to reason through retrieved examples and self-correct using reflection.

\subsection{Agentic Workflows in Software Engineering}

Agentic AI frameworks extend beyond static prompting by enabling LLMs to plan, act, and reflect~\cite{Ng2024AgenticWorkflow}. 
These workflows operationalize cognitive principles such as perception, memory, and reasoning through modular pipelines. 
Recent research in this area demonstrates that equipping LLMs with planning and reflection modules can significantly enhance reasoning quality, interpretability, and autonomy~\cite{Roychoudhury2025,Li2025,Wang2025AIAgentsWork}.

Ng’s \cite{Ng2024AgenticWorkflow} modular framework for agentic ai wokflow exemplifies this paradigm, showing how reasoning agents can use external tools, reflect on intermediate outputs, and refine their results iteratively. 
Acharya \textit{et al.}~\cite{Acharya2025Survey} survey these developments, noting that reflection allows models to identify weaknesses and self-correct in downstream tasks, effectively simulating human meta-cognition. 
Roychoudhury~\cite{Roychoudhury2025} further contextualizes this within software engineering, arguing that agentic AI represents a shift toward “autonomous but auditable” AI systems that can manage evolving codebases or perform software maintenance tasks with minimal supervision. 
Similarly, Li~\cite{Li2025} and Wang~\cite{Wang2025AIAgentsWork} propose role-based and collaborative multi-agent frameworks where agents handle subtasks such as retrieval, verification, and planning.

Reflection---a central component in our proposed workflow---has been explored in several prior works under different names. 
For example, the \textbf{Self-Refine} approach and related methods prompt LLMs to critique their own outputs before finalizing them. 
These reflective cycles, when combined with retrieval or tool usage (e.g., executing a code test, querying an API), lead to higher factual and logical accuracy~\cite{Ng2024AgenticWorkflow,Acharya2025Survey}. 
Our work aligns with this vision by combining retrieval grounding with structured self-reflection to identify whether a comment’s intent causally aligns with an observed code edit.

prior research has explored how Stack Overflow comments influence code edits~\cite{Soni2019,SHEIKHAEI2023111590,Tang2021}, how retrieval and zero-shot reasoning enhance LLM adaptability~\cite{Lewis2020,Parvez2021,Jiang2023,Ng2024AgenticWorkflow}, and how agentic frameworks introduce reflection and tool use~\cite{Acharya2025Survey,Roychoudhury2025,Li2025,Wang2025AIAgentsWork}. 
Our work synthesizes these directions by applying an \textbf{agentic retrieval-augmented reflective framework} to the problem of identifying comment-induced edits. 
Rather than fine-tuning or static classification, we combine retrieval grounding, reasoning, and reflection to emulate how human developers interpret and respond to code review comments, advancing toward interpretable and autonomous knowledge maintenance.

\section{RAG-Reflect: Retrieval Augmented LLMs with Reflection Workflows for Code Maintenance}
\label{sec:agentic-workflow}

\begin{figure}[t]
\centering
\includegraphics[width=0.9\linewidth]{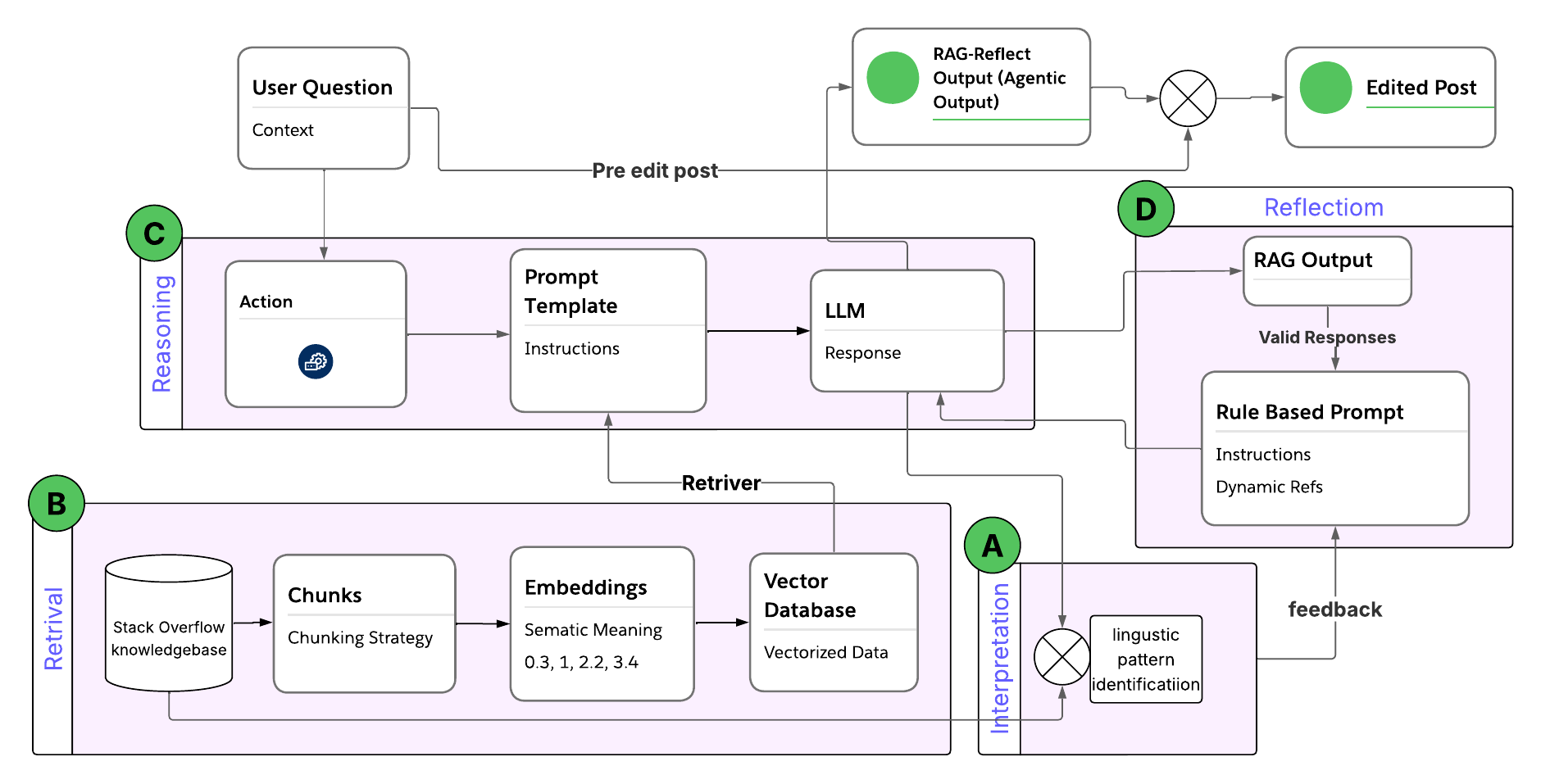}
\caption{Agentic retrieval-augmented workflow for comment-driven code maintenance. 
The pipeline comprises four sequential modules acting as autonomous agents.}
\Description{}
\label{fig:agentic_workflow}
\end{figure}

The RAG-Reflect framework consists of a one-time initialization phase \textbf{Interpretation} and a three-module runtime pipeline \textbf{Retrieval}, \textbf{Reasoning}, \textbf{Reflection}. These modules act as autonomous agents that collectively simulate the behavior of a developer analyzing feedback: recalling similar experiences, forming hypotheses, interpreting context, and validating conclusions. Figure~\ref{fig:agentic_workflow} illustrates this architecture, which integrates retrieval-augmented generation (RAG) with self-reflective validation to create a transparent, adaptable, and fine-tuning–free decision pipeline.

\noindent
The workflow begins with the \textbf{Interpretation} module operates as a one-time setup phase that analyzes the entire knowledge base to identify recurring patterns that distinguish valid from invalid comment-edit pairs. Unlike the other modules that execute per test instance, the Interpretation agent runs once during system initialization to extract generalized linguistic and structural cues from the training corpus.

Specifically, this module:
\begin{itemize}
    \item Analyzes the complete knowledge base (e.g., training set) to identify systematic patterns
    \item Discovers recurring linguistic markers, structural relationships, and semantic characteristics that differentiate valid from invalid pairs

    \item Generates a comprehensive set of validation rules that capture these patterns

\end{itemize}
\noindent
For example, the analysis might reveal that valid comments frequently contain imperative verbs ("add", "fix", "remove"), direct code references, and show high lexical overlap with edited code lines, while invalid comments often express gratitude, approval, or discuss unrelated topics. The output is a fixed rule-based prompt (as shown in Figure \ref{fig:reflection-prompt}) that encapsulates these discovered patterns and serves as a consistent validation template for all test instances.

\textbf{Retrieval} module, which serves as the contextual memory of the system. Instead of treating each input as an isolated case, the agent queries an indexed library of historical comment–edit pairs drawn from the SOUP corpus. Each record in this knowledge base — containing the comment, its corresponding code before and after an edit, and a binary validity label — is represented as a dense embedding using the MiniLM \footnote{url{https://huggingface.co/sentence-transformers/all-MiniLM-L6-v2}} sentence encoder and indexed via FAISS\footnote{url{https://github.com/facebookresearch/faiss}} for efficient vector search. Given a new query, the retrieval agent identifies the most semantically similar historical examples by computing cosine similarity between the query and stored embeddings. The top-$3$ matches  are selected as contextual exemplars. This mechanism allows the system to leverage “collective experience”: if the incoming comment resembles a past case that led to a valid code change, the model can infer analogous reasoning. For example, a comment such as “handle null input before calling the function” would likely retrieve prior instances where similar feedback led to edits introducing conditional checks. In essence, the retrieval module externalizes memory — grounding the model’s inference in concrete precedents without retraining.

Once contextual knowledge is gathered, the \textbf{Reasoning} module performs causal inference to determine whether the current comment plausibly motivated the code modification. A large language model (LLM), GPT-4o in our implementation, is prompted in a structured zero-shot setting that combines the current pair (comment, code before, code after) with the retrieved examples. The model is instructed to decide whether the comment likely triggered the observed edit, providing a concise justification linked to the code diff. This setup enables case-based reasoning: the model contrasts the present situation with retrieved analogues, reasoning by similarity rather than by memorized patterns. Empirically, this stage demonstrates the largest performance gain, as retrieval-based prompting provides contextual grounding that reduces hallucinations and improves causality  \cite{Panda2025AgenticRAG}. The reasoning agent therefore acts as the cognitive core of the system — a flexible decision-maker capable of learning from analogical context in real time.

The final component, \textbf{Reflection}, operates as a per-instance self-correction mechanism that validates each Reasoning output against the fixed rules generated during interpretation. For every test instance, the Reflection agent:

\begin{itemize}
    \item Receives the initial validity decision from the Reasoning module

    \item Applies the pre-generated validation rules to assess consistency

    \item Corrects the decision if rule violations are detected

\end{itemize}

This separation of concerns—where pattern discovery happens once during setup, and rule application happens per instance—ensures both efficiency and consistency. The Reflection module can quickly validate each prediction against established patterns without the computational overhead of re-analyzing the entire knowledge base.

\noindent
\subsection{Overall Pipeline\\}
The RAG-Reflect pipeline operates through a coordinated sequence of initialization and inference phases. During system initialization, the Interpretation module performs a one-time analysis of the entire knowledge base to extract recurring patterns and generate a fixed set of validation rules. For each test instance, the workflow then executes three sequential steps: first, the Retrieval agent identifies semantically similar historical examples to provide contextual grounding; second, the Reasoning agent analyzes the current comment-edit pair in light of these analogies to produce an initial validity decision; and finally, the Reflection agent applies the pre-established validation rules to critique and refine this decision, creating an efficient self-correcting loop that maintains pattern-guided validation without per-instance analysis overhead. This modular design separates expensive pattern discovery from instance-specific reasoning, ensuring both computational efficiency and consistent, well-grounded predictions across all test cases.

Together, these moodules — interpretation, retrieval, reasoning, and reflection — form a coherent perception–analysis–evaluation loop. The pipeline transforms raw Stack Overflow history into a continuous process of analogical learning and introspective validation. The the interpretation agent translates judgment into human-readable rationale, retrieval agent brings external memory; the reasoning agent synthesizes context into a causal judgment; and the reflection agent enforces logical and empirical constraints from the interpretation. This modular composition promotes transparency, adaptability, and accountability. Because each stage exposes its intermediate artifacts, the workflow can be inspected, debugged, and improved at the component level. For instance, researchers can analyze retrieved examples to assess memory coverage, study reasoning outputs to evaluate the quality of causal inference, or examine reflection logs to refine decision rules. Ultimately, the agentic paradigm converts VCP from a static classification task into an intelligent reasoning process—one that continuously learns, critiques, and justifies its own decisions while maintaining fine-tuned-level performance in an alternative approach.

\section{Experimental Setup}
\label{sec:evaluation}
This section discusses the procedure we use to evaluate the performance of our proposed framework for predicting valid comment-edit pairs.

\subsection{Research Questions}
Our empirical analysis is guided by the following research questions:

\begin{itemize}
    \item \textbf{RQ1: Overall Performance.}~How does \textbf{RAG-Reflect} perform in identifying valid comment-edit pairs?
    \item \textbf{RQ2: Ablation Study.}~How does each component contribute to the overall performance?
    \item \textbf{RQ3: Effectiveness of Different Prompt Engineering
Strategies.} How do different standalone prompt engineering strategies influence the effectiveness of our approach?
\end{itemize}



\subsection{Datasets}
\label{sec:datasets}
For the purpose of evaluation, we use two different datasets. The data collection process consists of the following steps. The first step is to collect all edit history associated to a SO post (i.e., either a question or an answer) and comments. The next step is to randomly sample comment-edit pairs since there are a large number of comment-edit pairs and it is not feasible to manually validate all of them. The final step is to categorize the selected pairs into valid and invalid categories with the help of human annotators. 

Creating such a dataset is a time-consuming operation. Thus, we collect the dataset published by Mai et~al.~\cite{Mai2025SOUP}, which consists of 5{,}000 manually annotated comment–edit pairs collected from Java answers with code examples. We refer to this as the \textit{soup-dataset}. Each pair includes the \texttt{code\_before}, \texttt{code\_after}, and \texttt{comment\_text} triplet, labeled as \textbf{Valid (1)} or \textbf{Invalid (0)} depending on whether the comment triggered the observed edit. Annotation followed a double-blind protocol with inter-rater agreement \cite{Cohen1968} $\kappa=0.795$. We adopt the original split of 4{,}000 training, 500 validation, and 500 test samples to ensure reproducibility of study results.

We create the second dataset by considering Python answer posts with code examples. We collect comment–edit pairs from Stack Overflow answers with examples between January~2022 and December~2024 using Stack Exchange Data Explorer API. We then apply the following criteria: (i) we select comments with at least one up-vote, (ii) edit happens after the comment as our goal is to predict the comment that triggers the edit (iii) if multiple edits occur after a comment, we pair the comment with the closest edit following the construction of the soup-dataset. The first two authors of the papers then independently annotate 1000 randomly sampled comment-edit pairs. Both authors have more than seven years of experience programming in Python and have used Stack Overflow for more than ten years. The authors consider the answer and the associated question to understand the problem and independently annotate the comment-edit pairs into valid and invalid categories. If there is a disagreement, the authors discuss with each-other to reach a conclusion. If we cannot reach any conclusion, we remove the comment-edit pair from our analysis. We compute Cohen’s kappa [9] to measure inter-rater agreement in classifying comment-edit pairs into valid and invalid categories. We obtain a kappa value of 0.81 prior to resolving the disagreements, which indicates a high level of agreement. At the end, we manually annotated 1{,}000 comment-edit-pairs and 143 of those pairs are identified as valid. We then constructed a Python-based VCP evaluation set by sampling 572 comment–code pairs, consisting of 143 valid and 429 invalid instances, preserving the original 1:3 valid–invalid distribution used in the prior study~\cite{Mai2025SOUP}. This dataset allows us to assess generalization beyond the Java domain. 

\usetikzlibrary{arrows.meta,positioning,calc,fit,shadows.blur}
\definecolor{cUser}{RGB}{230,230,230}     
\definecolor{cMem}{RGB}{173,202,255}      
\definecolor{cRAG}{RGB}{200,235,200}      
\definecolor{cReason}{RGB}{252,204,204}   
\definecolor{cInterp}{RGB}{210,230,255}   
\definecolor{cRefl}{RGB}{255,235,179}     
\definecolor{cOut}{RGB}{186,215,174}      
\tikzset{
  box/.style={rounded corners, draw=black!70, align=center, font=\scriptsize,
              minimum width=25mm, minimum height=8mm, inner sep=3pt,
              blur shadow},
  arrowF/.style={-Latex, very thick},
  arrowB/.style={-Latex, very thick, dashed},
  labeltiny/.style={font=\scriptsize, inner sep=1pt, fill=white, text=black}
}

\subsection{Evaluation Metrics}

We evaluate all models using three standard classification metrics—\textbf{Precision}, \textbf{Recall}, and \textbf{F1-score}—which jointly assess the accuracy and completeness of valid comment–edit prediction.

\textbf{Precision} measures how many of the comment–edit pairs predicted as valid are actually correct. It is defined as follows:
\[
\text{Precision} = \frac{TP}{TP + FP}
\]
where $TP$ and $FP$ denote true positives and false positives, respectively.

\textbf{Recall} quantifies the model’s ability to identify all true valid comment-edit pairs within the dataset. It is calculated as follows:
\[
\text{Recall} = \frac{TP}{TP + FN}
\]
where $FN$ represents false negatives—valid pairs that the model failed to detect.

\textbf{F1-score} provides a single harmonic mean between Precision and Recall, balancing precision-oriented and recall-oriented behavior:
\[
\text{F1} = \frac{2 \times \text{Precision} \times \text{Recall}}{\text{Precision} + \text{Recall}}
\]
  
For the agentic pipeline, we record both pre- and post-reflection values of these metrics to measure the effect of the reflection stage on overall decision accuracy and false-positive reduction.
\subsection{Compared Techniques}

For the purpose of evaluation and comparison of \textbf{RAG-Reflect} we consider the following techniques:

\begin{itemize}
    \item \textbf{Tang's Keyword matching technique:} Tang et al. \cite{Tang2021} considered keyword matching approach to detect the comments that can trigger an edit generation. They proposed a simple matching-based method to determine whether an edit is related
    to a comment. Specifically, their method matches a comment
    to an edit based on three heuristic rules: (1) the comment
    occurred before the edit; (2) the comment mentions a code
    term that gets added or removed from a code snippet within
    the edit; (3) the commenter and the editor are different users. Keyword based matching technique often fails to capture the semantic meaning or the intent of the comment while there are no keyword overlapping between the edit diff and the comment.

    \item \textbf{SOUP Technique:} Mai et al. \cite{Mai2025SOUP} proposed the valid comment edit pair identification (VCP) and automatic post update (APU) using the fine tuned LLM. They used 4000 of manually annotated data for the fine tuning using  the low-rank adaptation (LoRA) \cite{lora}. They run Tang’s method on their
    human annotated test set for evaluation and found their fined tuned model's performance outperforms Tang's approach.

    \item \textbf{Feature Based Techniques:} prior works used traditional machine learning models for comment classification task \cite{ml1, ml2, ml3}. Inspired by them we also extract some features from the training 4000 data and perform Logistic Regression, Random Forest, XGBoost algorithms for the VCP task. We began with a qualitative pass over 384 SOUP instances using a standard sample-size calculator \footnote{\url{https://www.calculator.net/sample-size-calculator.html}} to hypothesize salient cues (politeness vs.\ directive phrasing, action/correction verbs, code-like references, and code–comment lexical overlap). Guided by these observations, we engineered a compact but expressive set of 24 features spanning five categories: \emph{Lexical/Length}, \emph{Pragmatics}, \emph{Action/Error Cues}, \emph{Code-like/References}, and \emph{Structure/Overlap}. Table~\ref{tab:features} documents these features and extraction cues.
    We standardized all numeric features and trained a logistic regression (\textsc{LR}) model to obtain \emph{standardized} coefficients $\beta$ and their \emph{standard errors} (SE) via the usual $(X^\top W X)^{-1}$ estimator.

    \item \textbf{Agentic Technique (ours):}
To assess the autonomous reasoning capability of our system, we evaluate three progressively enhanced configurations of the proposed agentic framework. The first variant, \textbf{RAG (Retrieval-Augmented Generation)}, enriches reasoning with contextual grounding. It retrieves the top-$k$ ($k=3$) semantically similar comment–edit pairs from the SOUP knowledge base using MiniLM embeddings and FAISS indexing. These retrieved analogies act as real-world exemplars that guide the model in determining whether a given comment–edit pair reflects a valid causal relationship.  

The second configuration, \textbf{RAG + Reflection}, incorporates an additional verification layer. Here, the model’s RAG reasoning output is passed through a rule-based filter that operationalizes the linguistic cues discovered in Interpretation stage of our workflow—such as action-oriented verbs (\textit{fix, update, add}), gratitude tokens (\textit{thanks, works fine}), and lexical overlap thresholds between comment and modified code. This hybrid rule-augmented reasoning improves precision by eliminating false positives.

Finally, the \textbf{RAG-Reflect} configuration represents the complete agentic workflow. It integrates retrieval, reasoning, and reflection into a self-correcting feedback loop. The pipeline begins by retrieving relevant historical examples, reasoning over the comment–code pair in a zero-shot fashion, and then validating its decision against explicit linguistic, semantic, and temporal constraints. If inconsistencies are found—for example, a gratitude-only comment being classified as valid—the reflection module prompts the LLM to reassess its reasoning and adjust the output. This iterative reasoning–reflection process exemplifies agentic intelligence: the model autonomously grounds, reasons, and verifies its own judgment without any fine-tuning.

\medskip
\noindent
The design of the prompt templates for both the reasoning and reflection stages is critical to maintaining consistency, clarity, and logical rigor. Figure~\ref{fig:rag-reasoning-prompt} and \ref{fig:reflection-prompt} illustrates the visualization of these two prompt types used in RAG-Reflect.

\begin{figure}[t]
\centering
\begin{tcolorbox}[colframe=black!70, colback=white!98!black, arc=2mm, boxrule=0.6pt, width=0.95\columnwidth]
\footnotesize
\textcolor{blue!70!black}{\textbf{Role:}} You are a Stack Overflow moderator specializing in comment-driven code maintenance.\\[0.3em]
\textcolor{green!40!black}{\textbf{Task:}} Evaluate whether a comment justifies the corresponding code edit using both current and historical examples.\\[0.4em]
\textcolor{orange!70!black}{\textbf{Retrieved Similar Examples (Top-3):}}\\
\texttt{Example1: [Comment–Edit Pair]}\\
\texttt{Example2: [Comment–Edit Pair]}\\
\texttt{Example3: [Comment–Edit Pair]}\\[0.4em]
\textcolor{cyan!50!black}{\textbf{Context:}}\\
\texttt{Comment: \{comment\}}\\
\texttt{Code Before: \{code\_before\}}\\
\texttt{Code After: \{code\_after\}}\\[0.4em]
\textcolor{red!70!black}{\textbf{Instruction:}} Decide if the comment justifies the edit. Respond strictly with “\texttt{YES}” or “\texttt{NO}”.\\[0.3em]
\textcolor{purple!70!black}{\textbf{Expected Output:}} \texttt{YES / NO}
\end{tcolorbox}
\caption{Retrieval–Reasoning prompt used in RAG-Reflect. The model analyzes the comment–code pair with reference to the top-3 retrieved analogies.}
\label{fig:rag-reasoning-prompt}
\end{figure}

\begin{figure}[t]
\centering
\begin{tcolorbox}[colframe=black!70, colback=white!98!black, arc=2mm, boxrule=0.6pt, width=0.95\columnwidth]
\footnotesize
\textcolor{blue!70!black}{\textbf{Role:}} You are a peer reviewer and trained on stack overflow comment-edit mechanism. You are verifying whether an edit was directly motivated by a comment on Stack Overflow.\\[0.3em]
\textcolor{green!40!black}{\textbf{Decision Rules:}}\\
\textbf{Answer “YES” if all are true:}\\
-- The comment implies or requests a change addressed by the edit.\\
-- There is semantic or lexical overlap between the comment and changed code.\\
-- The comment appears before the edit and could have triggered it.\\[0.3em]
\textbf{Answer “NO” if any are true:}\\
-- The comment is gratitude-only or approval-only.\\
-- The comment discusses something unrelated to the code changes.\\
-- The edit fixes an issue not mentioned in the comment.\\
-- The comment occurs after the edit.\\[0.4em]
\textcolor{cyan!50!black}{\textbf{Input:}}\\
\texttt{code\_before: <original code>}\\
\texttt{code\_after: <modified code>}\\
\texttt{comment: <text of the comment>}\\[0.4em]
\textcolor{red!70!black}{\textbf{Task:}} Compare the code versions and comment. Output strictly one word: \texttt{YES} or \texttt{NO}.\\[0.3em]
\textcolor{purple!70!black}{\textbf{Expected Output:}} \texttt{YES / NO}
\end{tcolorbox}
\caption{Rule-based Reflection prompt. The model validates causal consistency using explicit linguistic and temporal rules.}
\label{fig:reflection-prompt}
\end{figure}

\medskip
Together, these prompts operationalize the full RAG-Reflect pipeline. The retrieval stage supplies external memory, the reasoning stage performs semantic causality assessment, and the reflection stage enforces logical consistency. This layered prompting strategy converts LLM-based classification into a human-like cognitive process—grounded, self-aware, and verifiable.

\end{itemize}

\begin{figure}[t]
\centering
\begin{tcolorbox}[
    colframe=blue!60!black,
    colback=blue!3,
    arc=2mm,
    boxrule=0.8pt,
    width=0.95\columnwidth]
\small

\textcolor{blue!70!black}{\textbf{Context:}}\\
You are verifying whether a code edit was directly motivated by a user comment.
Determine whether the comment justifies the changes made from 
\texttt{code\_before} to \texttt{code\_after}.

\textcolor{blue!70!black}{\textbf{Code Before:}}\\
\{\texttt{code\_before}\}

\textcolor{blue!70!black}{\textbf{Code After:}}\\
\{\texttt{code\_after}\}

\textcolor{blue!70!black}{\textbf{Comment:}}\\
\{\texttt{comment}\}

\textcolor{blue!70!black}{\textbf{Task:}}  
Does the comment justify the edit? Respond with one word.\\[0.3em]

\textbf{Expected Output:}
\begin{itemize}
    \item \texttt{valid} — the edit is justified by the comment
    \item \texttt{invalid} — the edit is not justified by the comment
\end{itemize}

\end{tcolorbox}
\caption{Zero-shot prompt template used for the Valid Comment--Edit Prediction (VCP) task.}
\label{fig:vcp-zeroshot}
\end{figure}

\begin{table}[t]
\centering
\caption{Engineered features with categories, definitions, and extraction cues.}
\label{tab:features}
\scriptsize
\begin{tabular}{p{2.2cm}p{3.0cm}p{3.8cm}p{3.6cm}}
\toprule
\textbf{Category} & \textbf{Feature} & \textbf{Definition / Rationale} & \textbf{Extraction Cue (example)} \\
\midrule
Lexical/Length &
\texttt{char\_count}, \texttt{word\_count}, \texttt{sentence\_count}, \texttt{avg\_sentence\_len}, \texttt{lexical\_diversity} &
Verbosity and specificity: valid comments are concise but content-dense. &
Tokenize sentences/words; unique/total tokens. \\ \hline

Pragmatics &
\texttt{has\_thanks}, \texttt{has\_please}, \texttt{has\_modal}, \texttt{num\_questions}, \texttt{is\_question}, \texttt{num\_exclaims} &
Politeness/acknowledgment vs.\ suggestion/hedging. Invalids often gratitude-only. &
Regex \\ \hline

Action/Error Cues &
\texttt{contains\_fix\_words}, \texttt{contains\_negative} &
Directive verbs and problem words signal actionable feedback. &
Regex \\ \hline

Code-like/Refs &
\texttt{has\_code\_ticks}, \texttt{has\_identifier\_style}, \texttt{has\_function\_pattern}, \texttt{num\_digits}, \texttt{code\_like\_ratio}, \texttt{ref\_tokens}, \texttt{avg\_token\_len} &
Concrete references to code entities increase validity likelihood. &
Backticks; camel/snake case \\ \hline

Structure/Overlap &
\texttt{loc\_before}, \texttt{num\_functions\_before}, \texttt{num\_comments\_in\_code}, \texttt{token\_overlap\_count}, \texttt{token\_overlap\_ratio} &
Complexity and code–comment semantic alignment. &
Line counts, function keywords; token overlap \\
\bottomrule
\end{tabular}
\end{table}


\paragraph{Large Language Models.}
We evaluate five representative LLMs listed in Table \ref{tab:model-summary} that capture diverse reasoning and code-understanding paradigms. \textbf{GPT-4o} serves as a high-performing general-purpose model with strong reasoning and multilingual capabilities. \textbf{Gemini~2.5~Flash} represents a fast, multimodal, instruction-following model optimized for low-latency inference. \textbf{Qwen-30B} provides a powerful open-source alternative with strong analytical and code comprehension abilities. In addition, we include code-specialized transformer model: \textbf{CodeLlama-13B} which is tuned for code synthesis, transformation, and reasoning tasks. All models are used in a \emph{training-free} configuration to ensure fairness and reproducibility across experiments. For comparability, we standardize decoding hyperparameters across all prompting strategies, using a temperature of~0.0 to eliminate stochastic variability.

\begin{table}[t]
\centering
\caption{Models evaluated in this study.}
\label{tab:model-summary}
\scriptsize
\begin{tabular}{lcccl}
\toprule
\textbf{Model} & \textbf{Type} & \textbf{Params} & \textbf{Access} & \textbf{Role in Study} \\
\midrule
GPT-4o & General LLM & N/A & API & Zero-/Few-shot, RAG-Reflect \\
Gemini~2.5~Flash & Multimodal LLM & N/A & API & Few-shot, RAG-Reflect \\
Qwen-30B & Reasoning LLM & 30B & API/Local & Zero-/Few-shot, CoT \\
Codellama 13B & Code-Specialized LLM & 30B & API/Local & Zero-/Few-shot, CoT \\
Logistic Regression & Baseline ML & -- & Local & Feature-based baseline \\
Random Forest & Baseline ML & -- & Local & Feature-based baseline \\
XGBoost & Baseline ML & -- & Local & Feature-based baseline \\
\bottomrule
\end{tabular}
\end{table}
\subsection{Implementation Details}
All experiments are conducted on the publicly available SOUP dataset~\cite{Mai2025SOUP}, containing 5{,}000 annotated comment–edit pairs (27\% valid, 73\% invalid). Following the original split, we use 4{,}000 pairs for training or retrieval indexing, 500 for validation, and 500 for final testing. The FAISS index is constructed from Sentence-BERT MiniLM embeddings of each $(\texttt{comment}, \texttt{code\_before}, \texttt{code\_after})$ triple, enabling sub-second semantic retrieval. For the reflection module, we implement deterministic rule checks derived from the training data provided by SOUP framework using LLM, we call it as an Interpretation agent. LLM outputs are collected via batched API calls and post-processed with regular expressions to extract binary labels. our replication package is available at github.

\section{Experimental Results}
\label{sec:results}

\subsection{RQ1 — Overall Performance}
\label{sec:rq1}

\begin{table*}[t]
\centering
\caption{Results of different techniques on VCP (Valid Comment–Edit Prediction).}
\label{tab:overall-performance}
\begin{tabular}{lccc|ccc}
\toprule
\multirow{2}{*}{Model} 
& \multicolumn{3}{c|}{Invalid Class } 
& \multicolumn{3}{c}{Valid Class } \\
\cmidrule{2-7}
& Precision & Recall & F1 
& Precision & Recall & F1 \\
\midrule
CodeLlama-13B (Zero-Shot)        & 0.76 & 0.55 & 0.64 & 0.27 & 0.50 & 0.35 \\
GPT-4o (Zero-Shot)               & 0.88 & 0.82 & 0.85 & 0.56 & 0.67 & 0.61 \\
Gemini 2.5 Flash (Zero-Shot)     & 0.92 & 0.76 & 0.83 & 0.53 & 0.81 & 0.64 \\
Qwen-Omni 30B  (Zero-Shot)       & 0.87 & 0.95 & 0.91 & 0.80 & 0.57 & 0.67 \\
Logistic Regression (feature-based)  & 0.84 & 0.68 & 0.75 & 0.39 & 0.61 & 0.48 \\
Random Forest  (feature-based)      & 0.76 & 0.95 & 0.84 & 0.46 & 0.13 & 0.21 \\
XGBoost   (feature-based)           & 0.79 & 0.89 & 0.84 & 0.47 & 0.29 & 0.36 \\
Tang's (matching-based)     & --- & --- & --- & 0.57 & 0.12 & 0.20 \\
SOUP      & --- & --- & --- & 0.80 & 0.74 & 0.77 \\
\textbf{RAG-Reflect (Ours)} 
                     & \textbf{0.92} & \textbf{0.95} & \textbf{0.94} 
                     & \textbf{0.81} & \textbf{0.74} & \textbf{0.78} \\
\bottomrule
\end{tabular}
\end{table*}

Table~\ref{tab:overall-performance} summarizes the result of our evaluation. Performance is reported separately for the invalid class and the valid class, since the task is inherently asymmetric: invalid pairs tend to be more common and often easier to detect, while valid pairs require deeper semantic understanding.

Across all systems, RAG-Reflect, our proposed method, achieves the strongest overall performance. For the invalid comment-edit pairs, RAG-Reflect obtains the highest precision (0.92), recall (0.95), and F1 score (0.94), indicating that it is highly reliable at filtering out mismatched comment–edit pairs. Notably, it exceeds the closest LLM competitor (e.g., GPT-4o Zero-Shot with an F1 score of 0.85) by a substantial margin. The Zero-Shot prompt template is in figure \ref{fig:vcp-zeroshot}.

Performance differences are more pronounced on the valid category, which is the more challenging category. Traditional feature-based models (Logistic Regression, Random Forest, XGBoost) show moderate recall but struggle with precision, leading to F1 scores below 0.50. Heuristic approaches such as Tang’s matching-based approach lagged significantly, reflecting the limitations of surface-level lexical cues. In contrast, modern LLMs demonstrate stronger semantic generalization: GPT-4o and Gemini 2.5 Flash achieve F1 scores of 0.61 and 0.64, respectively, while Qwen-Omni 30B leads among LLMs with an F1 of 0.67 specially for the valid class.

SOUP, the state-of-the-art technique specifically designed for predicting valid comment-edit pairs, performs competitively with an F1 of 0.77 on the valid class. However, RAG-Reflect surpasses all baselines and LLMs, achieving an F1 of 0.78. This improvement highlights the benefit of grounding the model through retrieval and structured reflection, enabling more accurate alignment between natural-language comments and code edits.


\subsection{RQ2: Ablation Study}
\label{rq2}

\begin{table}[t]
\centering
\caption{Ablation of RAG-Reflect components on the VCP task. We report precision (P), recall (R), and F1 for both invalid and valid pairs.}
\label{tab:rq2-ablation}
\begin{tabular}{lccc|ccc}
\toprule
\multirow{2}{*}{Variant} 
& \multicolumn{3}{c|}{Invalid (0)} 
& \multicolumn{3}{c}{Valid (1)} \\
\cmidrule{2-7}
& P & R & F1 & P & R & F1 \\
\midrule
RAG-only 
  & 0.92  & 0.69  & 0.79  & 0.47  & 0.83  & 0.60 \\
Reflection-only 
  & 0.91  & 0.91  & 0.91  & 0.73  & 0.74  & 0.73 \\
\textbf{Full RAG-Reflect} 
  & \textbf{0.92} & \textbf{0.95} & \textbf{0.94} 
  & \textbf{0.81} & \textbf{0.74} & \textbf{0.78} \\
\bottomrule
\end{tabular}
\end{table}

To assess the impact of each component in our agentic workflow, we performed a systematic ablation study. We consider three different configurations for the purpose of evaluation: (1) \textit{RAG-only}, where retrieved examples are provided but without self-reflection; (2) \textit{Reflection-only}, where the model critiques and revises its own initial prediction without external retrieval; and (3) the full \textbf{RAG-Reflect} workflow. This setup allows us to isolate the effect of retrieval, reflection, and their combined interaction. Table \ref{tab:rq2-ablation} populated the performance of individual components and highlight their contributions.

\paragraph{Standalone Behavior of Each Stage.}
Since direct generation (GPT-4o Zero-Shot) serves as the strongest baseline, achieving a valid F1 of 0.61. This confirms that a powerful frontier model can capture some semantic alignments between comments and edits but remains limited in recall (0.67), frequently missing subtle comment signals. In our workflow \textbf{RAG-only} demonstrates complementary behavior: access to retrieved examples substantially boosts valid recall (0.83) but introduces significant noise, lowering precision (0.47). This illustrates that retrieval alone is insufficient—without guided reasoning, the model tends to overgeneralize and classify many borderline cases as valid. \textbf{Reflection-only} shows the opposite pattern. By inspecting and refining its own prediction, the model becomes more conservative and accurate, producing high invalid F1 (0.91) and a balanced valid F1 (0.73). Reflection stabilizes judgments and reduces misclassifications, indicating that structured self-critique helps the model detect inconsistencies in its initial decision.

\paragraph{Effectiveness of the Full Workflow.}
The full \textbf{RAG-Reflect} workflow outperforms all variants, achieving the highest invalid F1 (0.94) and the highest valid F1 (0.78). These gains emerge because retrieval and reflection provide complementary strengths: retrieval injects grounded context—such as similar past comment–edit examples—while reflection filters and aligns this information with the true intent of the comment. RAG-only improves recall but harms precision; reflection-only improves precision but loses contextual breadth; only their combination yields a solution that is simultaneously accurate, grounded, and robust. The full workflow demonstrates that VCP prediction is not a simple classification task but a multi-step reasoning process requiring both external knowledge and iterative verification. In summary, the ablation results validate that each component contributes meaningfully, and their integration produces the most reliable and generalizable performance for comment-driven post understanding.

\subsection{RQ3: Effectiveness of Different Prompt Engineering
Strategies.}
\label{sec:rq3}

\begin{table*}[t]
\centering
\caption{Comparison of Prompt Engineering Strategies for the VCP Task. We report F1-scores for the valid class (edit-worthy comments), along with invalid class results. RAG-Reflect is shown for reference.}
\label{tab:rq3-prompts}
\begin{tabular}{lcccc}
\toprule
\textbf{Model} & \textbf{Prompting Strategy} & \textbf{Invalid F1} & \textbf{Valid F1} \\
\midrule
GPT-4o            & Zero-Shot           & 0.85 & 0.61 \\
CodeLlama-13B     & Zero-Shot           & 0.64 & 0.35  \\
Gemini 2.5 Flash  & Zero-Shot           & 0.83 & 0.64  \\
Qwen-30B          & Zero-Shot           & 0.91 & 0.67  \\
\midrule
GPT-4o            & 2-shot              & 0.87 & 0.67  \\
CodeLlama-13B     & 2-shot              & 0.71 & 0.24  \\
Gemini 2.5 Flash  & 2-shot              & 0.76 & 0.60  \\
Qwen-30B          & 2-shot              & 0.86 & 0.63  \\
\midrule
GPT-4o            & 2-shot + CoT        & 0.84 & 0.64  \\
CodeLlama-13B     & 2-shot + CoT        & 0.19 & 0.42  \\
Gemini 2.5 Flash  & 2-shot + CoT        & 0.79 & 0.62  \\
Qwen-30B          & 2-shot + CoT        & 0.83 & 0.56  \\
\midrule
\textbf{RAG-Reflect (ours)} & N/A (agentic workflow) 
 & \textbf{0.936} & \textbf{0.777}  \\
\bottomrule
\end{tabular}
\end{table*}

\begin{figure*}[t]
\centering
\begin{tikzpicture}
\begin{axis}[
    ybar,
    bar width=5pt,
    width=0.98\linewidth,
    height=0.35\linewidth,
    ymin=0,
    ymax=1.0,
    ylabel={F1-score},
    symbolic x coords={
        GPT4o-ZS, CL13B-ZS, Gemini-ZS, Qwen-ZS,
        GPT4o-2S, CL13B-2S, Gemini-2S, Qwen-2S,
        GPT4o-CoT, CL13B-CoT, Gemini-CoT, Qwen-CoT,
        RAG-Reflect
    },
    xtick=data,
    xticklabel style={rotate=60, anchor=east},
    legend style={at={(0.5,1.15)},anchor=south,legend columns=2},
    ymajorgrids=true,
    grid style=dashed,
]

\addplot[fill=blue!60] coordinates {
    (GPT4o-ZS,0.85)
    (CL13B-ZS,0.64)
    (Gemini-ZS,0.83)
    (Qwen-ZS,0.91)

    (GPT4o-2S,0.87)
    (CL13B-2S,0.71)
    (Gemini-2S,0.76)
    (Qwen-2S,0.86)

    (GPT4o-CoT,0.84)
    (CL13B-CoT,0.19)
    (Gemini-CoT,0.79)
    (Qwen-CoT,0.83)

    (RAG-Reflect,0.936)
};

\addplot[fill=orange!70] coordinates {
    (GPT4o-ZS,0.61)
    (CL13B-ZS,0.35)
    (Gemini-ZS,0.64)
    (Qwen-ZS,0.67)

    (GPT4o-2S,0.67)
    (CL13B-2S,0.24)
    (Gemini-2S,0.60)
    (Qwen-2S,0.63)

    (GPT4o-CoT,0.64)
    (CL13B-CoT,0.42)
    (Gemini-CoT,0.62)
    (Qwen-CoT,0.56)

    (RAG-Reflect,0.777)
};

\legend{Invalid F1, Valid F1}

\end{axis}
\end{tikzpicture}
\caption{Comparison of Invalid and Valid F1-scores across multiple prompting strategies (Zero-Shot, 2-shot, 2-shot+chain-of-thought) and models, with RAG-Reflect as the final reference. RAG-Reflect demonstrates the best balanced performance across both classes.}
\label{fig:rq3-bar-full}
\end{figure*}
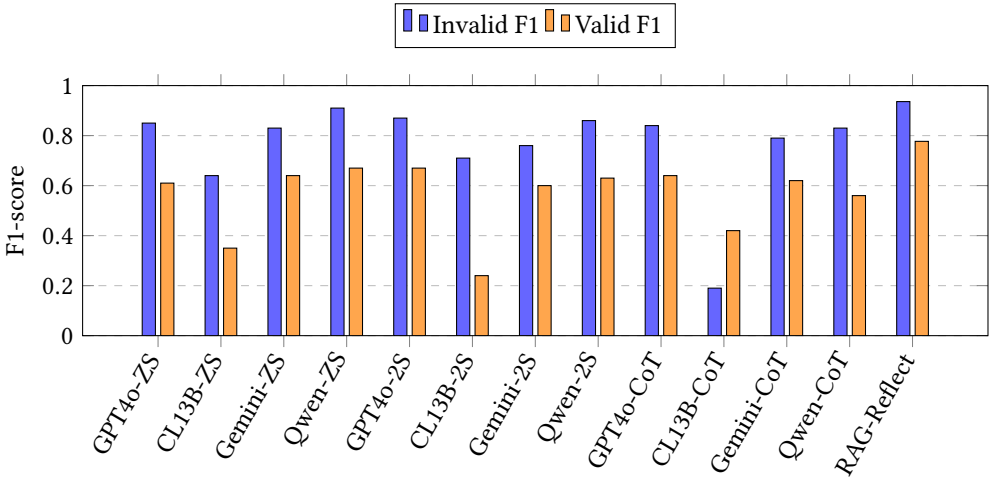

To understand whether prompt engineering alone can close the gap between generic LLM inference and our agentic workflow, we evaluated Zero-Shot prompting, two-shot prompting, and two-shot prompting with chain-of-thought (CoT) across four competitive models: GPT-4o, Gemini~2.5 Flash, CodeLlama-13B, and Qwen-30B. These strategies represent the most commonly-used prompting techniques in current LLM-based software engineering tools, as they require no fine-tuning and impose minimal computational overhead. Across all models, we observe substantial performance variability: Zero-Shot prompting produces a wide spread of valid-class F1 scores (0.35–0.67), two-shot prompting offers mild improvements for some models, and two-shot+CoT often increases recall but destabilizes precision, particularly for smaller models like CodeLlama-13B. The results show that prompting alone does not provide consistent or reliable control over the model’s ability to distinguish actionable comments from irrelevant ones—especially in scenarios requiring fine-grained semantic grounding.

A cross-strategy comparison further reveals fundamental limitations. While in-context examples help models like GPT-4o and Qwen-30B better recognize valid comments, they do not offer the contextual retrieval necessary to resolve ambiguous or implicit comment cues. Similarly, CoT encourages the model to reason more extensively, but without external evidence it often amplifies hallucinated justifications, harming precision. These behaviors stand in contrast to RAG-Reflect, which consistently outperforms all prompting variants with a valid F1 of 0.78 and an invalid F1 of 0.94. Unlike prompting-based methods, RAG-Reflect does not depend on handcrafted examples or verbal reasoning alone; instead, it integrates retrieval-grounded evidence with reflective self-critique to reduce both overgeneralization (a common failure of RAG-only baselines) and shallow heuristics (typical in Zero-Shot models). Overall, the findings indicate that prompt engineering—while helpful—is insufficient for reliable VCP prediction, and that structured agentic workflows provide the robustness and consistency required for high-stakes code maintenance tasks.

\section{Discussion}
\label{sec:discussion}

This section reflects on three broader dimensions of our study: (1) the generalizability of RAG-Reflect across programming languages, (2) a manual qualitative investigation of failure modes in the VCP task, and (3) the correctness and reliability of RAG-Reflect when applied to the Automatic Post Updating (APU) task. Together, these analyses provide deeper insight into the strengths and limitations of our agentic workflow and inform avenues for future research.

\subsection{Cross-Language Generalizability via the PyVCP Dataset}
To assess whether RAG-Reflect generalizes beyond Java-based SOUP data, we evaluated the full workflow on the newly introduced \textbf{PyVCP} corpus, a Python-specific comment--edit dataset containing more than 500 manually curated pairs. The results demonstrate that our method maintains strong discriminative power even in a different programming ecosystem. As shown in the final confusion matrix, RAG-Reflect achieves an F1-score of~0.91 for invalid pairs and~0.68 for valid pairs, with high precision for both classes (0.87 and 0.79, respectively). Notably, the model preserves excellent recall for invalid cases (0.95), suggesting strong robustness in identifying comments that \emph{do not} correspond to an actionable edit. While recall for valid cases decreases to~0.59, this behavior mirrors the expected cross-language difficulty—Python edits tend to be more structurally subtle, involve dynamic-typing considerations, and are less code-referential than Java edits.

These results highlight two important insights. First, the agentic retrieval–reasoning–reflection pipeline remains effective across languages: the high precision for valid pairs indicates that the model rarely hallucinated or over-predicted positive cases, which is critical for preventing incorrect automatic edits in practice. Second, the reduced recall for Python valid pairs points to a language-specific challenge: Python comments often rely on implicit conventions (“use list comprehension,” “avoid mutable default arguments,” “follow PEP8 style”), which may not surface strong lexical anchors or structural cues for retrieval. Despite this, RAG-Reflect still delivers balanced performance comparable to the Java results, confirming that the workflow generalizes across programming ecosystems without fine-tuning.

From a broader perspective, these findings suggest that the \emph{linguistic signals of actionable intent}—directive verbs, problem-identification phrases, and code-referential markers—are largely language-agnostic, even if the syntactic edits differ. This reinforces the viability of agentic VCP models in multilingual Q\&A environments. Future work may further improve cross-language recall by integrating language-specific retrieval indices, AST-level semantic checks, or syntax-aware reflection rules tailored to dynamically typed languages like Python.

\subsection{Failure Case Analysis of the RAG-Reflect Model on VCP task}

\begin{figure}[t]
\centering
\includegraphics[width=1\linewidth]{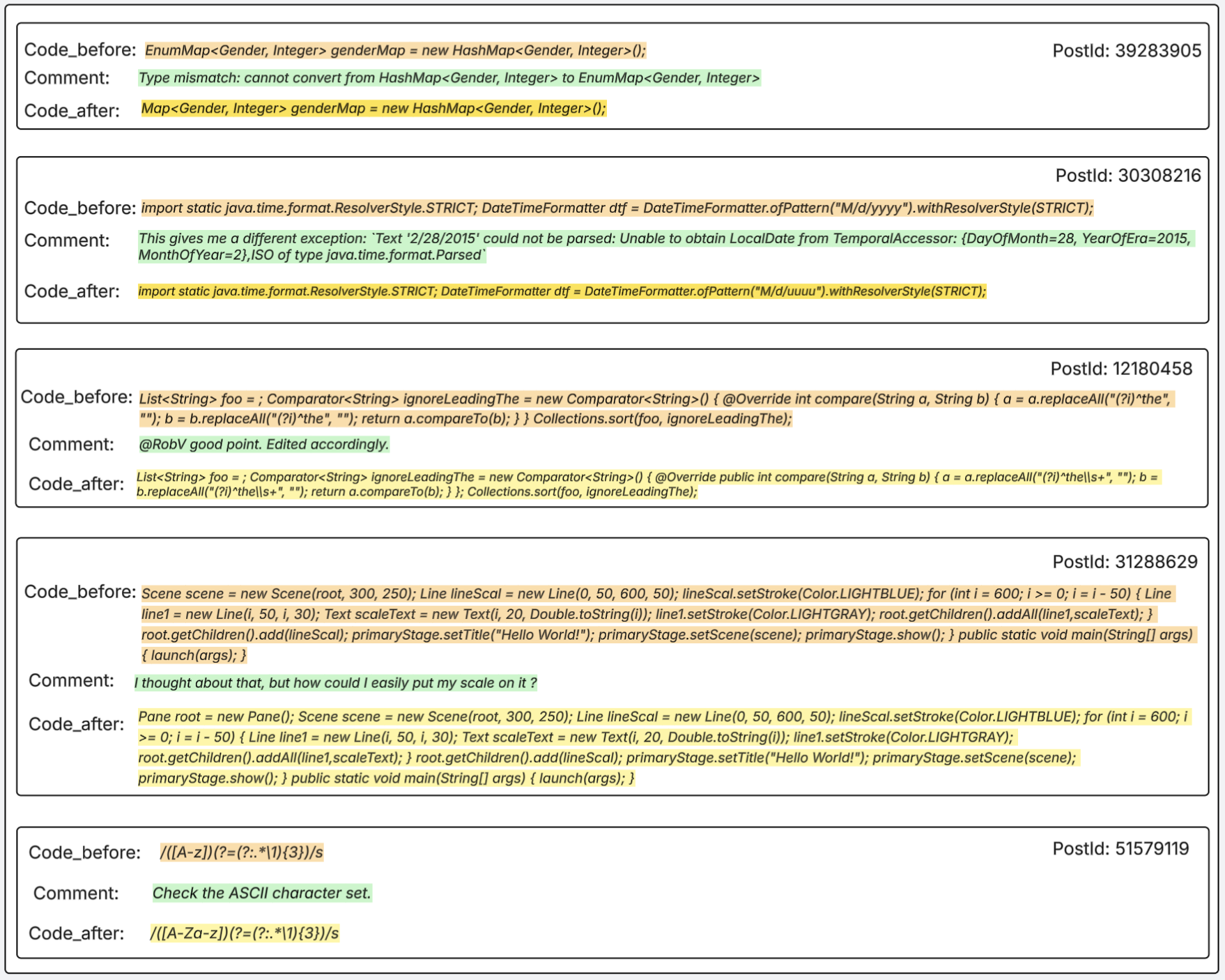}
\caption{Failure Cases analysis of RAG-Reflect}
\Description{}
\label{fig:failure cases}
\end{figure}

To understand the limitations of our RAG-Reflect classifier, we conducted an error analysis focusing specifically on the \textbf{true negative cases}---instances where the ground truth label was \texttt{YES}, but the model incorrectly predicted \texttt{NO}. These represent the most critical misclassifications because the model fails to recognize valid comment--edit alignment. Our analysis identifies five recurrent failure categories that reveal structural weaknesses in semantic reasoning, retrieval quality, and domain comprehension.

\subsubsection{Semantic Mismatch Between Comment and Code Change}
A common failure occurs when the comment and the code modification appear lexically related but address different underlying issues. In such scenarios, the model fails to recognize that the patch fully implements the developer’s intent. \textbf{Example (PostId 6923710).} The comment highlights a ``Type mismatch'' involving \texttt{EnumMap} and \texttt{HashMap}. The edit corrects the declared variable type, directly resolving the issue. Despite this clear alignment, the model predicts \texttt{NO}, revealing its tendency to rely on surface-level token cues rather than deeper semantic relationships between type systems and the applied fix.

\subsubsection{Trivial or Single-Line Fixes Misclassified as Non-Fixes} Simple syntactic edits---such as adjusting parentheses, replacing operators, or correcting date formats---are frequently overlooked. Because these changes affect only a few lines, the model often fails to interpret them as substantive repairs. \textbf{Example (PostId 30308216).} The comment indicates a date parsing problem. The patch updates the format from \texttt{``M/d/yyyy''} to \texttt{``M/d/uuuu''}. Although the fix directly corresponds to the comment, the model predicts \texttt{NO}, indicating reduced sensitivity to minimal but semantically meaningful changes.

\subsubsection{Generic or Non-Directive Comments}
Comments that contain acknowledgments or vague phrasing provide little explicit instruction for the classifier. Although the associated edits implement the intended correction, the absence of technical detail frequently leads the model to infer that no real fix occurred. \textbf{Example (PostId 12180458).} The comment states ``Edited accordingly,'' while the patch adds a missing \texttt{\};} to complete a comparator declaration. Despite the precise alignment between comment and edit, the classifier predicts \texttt{NO}, illustrating its difficulty handling non-descriptive, conversational comments.

\subsubsection{Multi-Line or Structural Changes Without Clear Local Cues}
Refactorings or structural code adjustments---such as modifying layout containers, adding wrapper elements, or introducing configuration annotations---require global reasoning about program behavior. The model struggles when the comment describes the issue at a high level and the fix spans multiple statements. \textbf{Example (PostId 31288629).} The comment asks how to ``put the scale on it'' in a JavaFX \texttt{Scene}. The patch introduces a new \texttt{Pane} and restructures the scene graph. The classifier predicts \texttt{NO} because the relationship between the abstract intent and the structural modification is not captured by simple lexical matching.

\subsubsection{Domain-Specific Fixes Requiring Deeper Technical Knowledge}
Misclassifications also arise when interpreting fixes that require knowledge of language-specific conventions, API rules, or regex semantics. The model lacks sufficient domain expertise to accurately judge whether the edit implements the correction described in the comment. \textbf{Example (PostId 51579119).} The comment notes that the regex \texttt{[A-z]} is invalid due to ASCII ordering. The patch updates the pattern to \texttt{[A-Za-z]}, which correctly captures alphabetic characters. Despite the exact semantic correspondence, the model predicts \texttt{NO}, revealing deficiencies in domain-level reasoning.

\subsection{Evaluation of Automatic post Update} 

To assess the functional behavior of RAG-Reflect in program transformation, we test \textbf{RAG-Reflect} and extend in a way to generate the automatic post update. Such an extension reveals the extensibility of our workflow. We provided the previous version of post and the valid comment predicted by the RAG-Reflect. We analyzed all examples in which our  agentic workflow generated code differed from the reference code. Because APU aims to emulate developer-intended edits expressed through natural-language comments, our evaluation emphasizes \textit{semantic and structural alignment} over raw string matching \cite{lin2004rouge, Ren2020CodeBLEU}. The analysis revealed five recurring correctness patterns, each influenced by how the model interprets the guiding comment.

\paragraph{1. Formatting and Layout-Only Differences}  
In many cases, the model correctly understands the comment and applies the intended transformation, but outputs a reformatted version of the reference code. Since the underlying semantics remain unchanged, the discrepancy originates from formatting variation rather than misunderstanding the comment. This style of divergence harms exact-match scores \cite{chen2018codexm}, but has no functional impact. \textbf{Example (PostId 39283905).} The comment describes how to populate a matrix with random integers. RAG-Reflect generates a structured, multi-line loop that faithfully implements the comment's intent, while the reference uses a compact one-line loop. The \texttt{exact match score} is 0.0 due to formatting, but BLEU is 1.0 because all operations required by the comment are preserved.

\paragraph{2. Syntax or Identifier Corrections Guided by Comment Semantics}  
When comments reference broken behavior, RAG-Reflect often ``over-corrects'' syntactic flaws not explicitly mentioned but logically related. These cases show that the LLM leverages both retrieved examples and comment semantics to infer required fixes \cite{yin2018learning}. \textbf{Example (PostId 41343549).}  
The comment critiques the window-disposal logic. The ground truth retains a placeholder reference (\texttt{nameofyourframe}), but RAG-Reflect updates it to \texttt{this}, reflecting a valid inference about the intended UI context. Although the comment does not explicitly request this substitution, the model makes a reasonable correction based on typical Java patterns. BLEU is 0.83, but the \texttt{exact match} is 0.

\paragraph{3. Completion of Missing Initialization Inferred from Comment Intent}  
Some comments describe behavior requiring variables or collections to be initialized. When the reference snippet contains placeholders, RAG-Reflect inserts plausible defaults. \textbf{Example (PostId 12180458).}  
The comment discusses operations on a list, but the reference code leaves the list uninitialized.  
The model correctly infers from the comment that the list must be instantiated, though this behavior deviates from the reference. BLEU remains moderate (0.68), signaling partial structural alignment.

\paragraph{4. Structural Boilerplate Completion Triggered by Comment Context}  
When comments refer to behavior requiring full class structure (e.g., threading, callbacks, or life cycle functions), RAG-Reflect often expands minimal reference code with standard boilerplate implementations. \textbf{Example (PostId 16922198).}  
The comment describes a threaded execution pattern. The reference provides only a skeleton, while the LLM produces a complete and logically coherent thread structure. Here BLEU < 1.0 because the generated structure satisfies the behavior implied by the comment, even though it diverges textually from the reference code.

\paragraph{5. Hallucinated or Semantically Irrelevant Code}  
Some failures arise when the model incorrectly interprets the comment, retrieves misleading contextual examples, or generates code unrelated to the intended edit. This reflects a retrieval–generation mismatch \cite{lewis2020retrieval}. \textbf{Example (PostId 43407388).}  
The comment concerns adding user information, but RAG-Reflect hallucinates Firebase event-listener logic unrelated to the expected method behavior. BLEU drops near zero (0.08), confirming that both lexical and structural similarity collapse in these comment-misinterpretation cases.

\paragraph{Metric-Based Performance Summary}  
To quantify the functional accuracy of RAG-Reflect in automatic post update, we report the following corpus-level scores:

\begin{itemize}
  \item \textbf{Exact Match Accuracy (em\_score):} 9.3\%
  \item \textbf{BLEU-4 with add-one smoothing (bleu4\_addone):} 0.71
\end{itemize}

Our manual inspection of the generated patches further confirms this mismatch. While a subset of outputs reveal genuine misinterpretations—typically cases where the model overfits to retrieved examples or infers an unintended control-flow transformation—the majority of non-matching updates are nevertheless semantically aligned with the ground truth. In many instances, RAG-Reflect reproduces the intended logic, preserves key data flow and control-flow patterns, and satisfies the constraints expressed by the comment, despite diverging from the reference in surface form.

\begin{figure}
\centering
\begin{tcolorbox}[
    colback=gray!5,
    colframe=gray!60,
    title=\textbf{APU Prompt Template},
    fonttitle=\bfseries,
    arc=2mm,
    boxrule=0.8pt,
    left=4pt,
    right=4pt,
    top=6pt,
    bottom=6pt
]
\textbf{Role:} You are a stack overflow post editor. You are expert at proposing edit based on the triggering comments on stack overflow.\\
\textbf{Task:} Update a Stack Overflow code snippet according to a reviewer comment.

\medskip
\noindent\textbf{Inputs:}
\begin{itemize}
    \item \texttt{code\_before}: \textcolor{blue!70}{\texttt{<ORIGINAL\_CODE\_SNIPPET>}}
    \item \texttt{comment}: \textcolor{blue!70}{\texttt{<USER\_COMMENT>}}
\end{itemize}

\medskip
\noindent\textbf{Instructions:}
\begin{enumerate}
    \item Modify the code \emph{only} to address the issue stated in the comment.
    \item Do not introduce refactoring, stylistic edits, or improvements not requested.
    \item Preserve the structure and formatting of the original snippet whenever possible.
    \item Return \textbf{only the updated code snippet}, with no explanation or prose.
\end{enumerate}

\medskip
\noindent\textbf{Output:}  
\textcolor{blue!80}{\texttt{<UPDATED\_CODE\_AFTER>}}
\end{tcolorbox}
\label{apu_promptx}
\caption{prompt template used for the Automatic Post Update (APU) task (an extension to our RAG-Reflect).}
\end{figure}

\subsection{Implications for Future Agentic Systems}
These three analyses reveal that RAG-Reflect’s strengths lie in (1) its robustness across languages, (2) its ability to reason about comment intent, and (3) its capacity to generate semantically meaningful code updates. At the same time, they highlight opportunities for integrating more rigorous semantic validation, improved intent grounding, and multi-agent collaboration. As agentic AI becomes increasingly central to software engineering~\cite{Acharya2025Survey,Ng2024AgenticWorkflow,Roychoudhury2025Trust}, workflows like RAG-Reflect can serve as foundational building blocks for autonomous code maintenance systems on large-scale platforms such as Stack Overflow.

\section{Threats to Validity}
\label{sec:threats_to_validity}
This section discusses threats to the validity of our research.

\subsection{External Validity}
Threats to external validity refer to the generalizability of our findings. We evaluate our proposed framework using code examples written in Java and Python. One can argue that the results may not generalize to other programming languages. Our selection of the languages is based on the fact that these are popular programming languages. We would like to point to the fact that our technique does not depend on any specific programming language or community-question answering website. Thus, our results should largely carry forward.
For the purpose of evaluating automatic post update task, we only considered the edits that were triggered by comments. However, it may be case edits were performed by the author of the post or external users without introducing any triggering comments. Although we do not consider such edits in this study, we can reuse the prompt template to recommend edits in those cases.

\subsection{Internal Validity}
Threats to internal validity refer to potential biases or errors in our research methodology. In this study, we conduct a manual study to categorize comment-edit pairs into valid and invalid categories. Such a manual study can introduce bias in the manually curated dataset. To avoid the bias, each comment edit-pair was independently evaluated by two different annotators. The level of inter-rater agreement of our qualitative study is also high.

To determine whether the recommended edits match to the actual edits performed by the SO user, we consider the Exact Match (EM) and BLEU as evaluation metrics. However, prior work has shown that metrics like CodeBLEU and BLEU scores are not reliable indicators of functional correctness. It may be the case that the generated edits are not textually similar to the edits performed by the users but they can be semantically equivalent. Since most code examples in SO are incomplete, it is difficult to execute the code and determine their functional correctness. To compensate this limitation, we manually validate the correctness of the recommended edits for automatic post update test cases.

\section{Conclusion}
\label{conclusion}
This paper presented RAG-Reflect, an agentic retrieval-augmented and self-reflective framework for Valid Comment–Edit Prediction (VCP) and its downstream application to Automatic Post Update (APU). For RQ1, our evaluation showed that RAG-Reflect consistently outperforms traditional feature-based models, embedding baselines, and strong zero-shot LLMs, achieving an F1 of 0.78 on valid cases and 0.94 on invalid ones, slightly surpassing SOUP while offering more stable behavior across examples. Through RQ2, our ablation analysis revealed that retrieval and reflection each contribute complementary strengths—retrieval broadens contextual understanding while reflection stabilizes reasoning—but neither is sufficient on its own. Their combination delivers the best precision–recall balance, demonstrating that VCP benefits from a multi-step reasoning workflow rather than a single forward pass. For RQ3, we observed that prompt-engineering strategies such as zero-shot, few-shot, and chain-of-thought yield inconsistent and model-dependent improvements, whereas RAG-Reflect provides dependable gains by grounding predictions in retrieved evidence and structured self-critique. Our APU study further showed that, the model often produces semantically appropriate edits, reflected by higher BLEU scores (0.71), with most discrepancies stemming from formatting or stylistic divergence rather than misunderstanding. Overall, RAG-Reflect demonstrates the potential of agentic LLM workflows to improve both the reliability of actionable comment identification and the quality of automated code updates, providing a clear, modular foundation for future research and practical integration.

\bibliographystyle{ACM-Reference-Format}
\bibliography{references}

\end{document}